%% file: main.tex
\documentclass[sigconf,balance=false]{acmart}

\input{utils/packages}

\input{utils/custom}

\input{utils/variables.tex}

\begin{document}

\date{}

\title[How Secure Messaging Apps Leak Sensitive Data to Push Notification Services]{The Medium is the Message: How Secure Messaging Apps Leak Sensitive Data to Push Notification Services}

\input{utils/authors}
\input{paper/abstract}


\maketitle

\input{index}
\begin{acks}
This work was supported by the U.S. National Science Foundation under grant CCF-2217771, the Center for Long-Term Cybersecurity (CLTC) at U.C. Berkeley, the KACST-UCB Center of Excellence for Secure Computing, an NSERC Discovery Grant, and a grant from the Silicon Valley Community Foundation. We would especially like to thank Chris Hoofnagle and Refjoh\"urs Lykkewe.
\end{acks}

\bibliographystyle{ACM-Reference-Format}
\bibliography{references}

\input{paper/appendix}

\end{document}

%% file: utils/packages.tex
\usepackage{popets}

\usepackage{tikz}
\usepackage{amsmath}

\usepackage{multirow}
\usepackage[normalem]{ulem}
\useunder{\uline}{\ul}{}
\usepackage{pdflscape}
\usepackage{lscape}
\usepackage{graphicx}
\usepackage{tabularx}

\usepackage{booktabs} 

\usepackage{wasysym}

\usepackage{breakurl}

\usepackage{listings}

%% file: utils/custom.tex
\setcopyright{popets}
\copyrightyear{YYYY}

\acmYear{YYYY}
\acmVolume{YYYY}
\acmNumber{X}
\acmDOI{XXXXXXX.XXXXXXX}
\acmISBN{}
\acmConference{Proceedings on Privacy Enhancing Technologies}
\def\BibTeX{{\rm B\kern-.05em{\sc i\kern-.025em b}\kern-.08em
    T\kern-.1667em\lower.7ex\hbox{E}\kern-.125emX}}

\newcommand{\mysubsubsection}[1]{\paragraph{\textbf{#1.}}}

%% file: utils/variables.tex
\newcommand{\selectedIMApps}{24}

\newcommand{\apps}{21}

\newcommand{\FCMApps}{20}

\newcommand{\leakedAny}{11}

\newcommand{\leakedDeviceIDs}{3}

\newcommand{\leakedUserIDs}{10}

\newcommand{\leakedNames}{7}

\newcommand{\leakedPhones}{2}

\newcommand{\leakedContents}{4}
\newcommand{\notLeakedContents}{7}

\newcommand{\mitigationApps}{16}

\newcommand{\completeMitigationApps}{9}

\newcommand{\incompleteMitigationApps}{7}

\newcommand{\etoeApps}{8}

\newcommand{\etoeCompleteApps}{4}

\newcommand{\etoePartialApps}{4}

\newcommand{\etoeLeakedUserIDs}{3}

\newcommand{\etoeLeakedNames}{3}

\newcommand{\ptosApps}{8}

\newcommand{\ptosCompleteApps}{5}

\newcommand{\ptosPartialApps}{3}

\newcommand{\ptosLeakedDeviceIDs}{2}

\newcommand{\ptosLeakedUserIDs}{3}

\newcommand{\ptosLeakedPhones}{2}


%% file: utils/authors.tex

\author{Nikita Samarin,$^{1,2}$ Alex Sanchez,$^1$ Trinity Chung,$^1$ Akshay Dan Bhavish Juleemun,$^1$ Conor Gilsenan,$^1$ Nick Merrill,$^1$ Joel Reardon,$^3$ and Serge Egelman$^{1,2}$}
\email{{nsamarin, alexso, trinityc, adbjuleemun, conorgilsenan, ffff, egelman}@berkeley.edu}\email{joel.reardon@ucalgary.ca}
\affiliation{%
\institution{$^1$University of California, Berkeley; $^2$International Computer Science Institute (ICSI); $^3$University of Calgary}
\country{} 
}

\renewcommand{\shortauthors}{N. Samarin et al.}

%% file: paper/abstract.tex
\begin{abstract}
Like most modern software, secure messaging apps rely on third-party components to implement important app functionality. Although this practice reduces engineering costs, it also introduces the risk of inadvertent privacy breaches due to misconfiguration errors or incomplete documentation. Our research investigated secure messaging apps' usage of Google's Firebase Cloud Messaging (FCM) service to send push notifications to Android devices. We analyzed \apps{} popular secure messaging apps from the Google Play Store to determine what personal information these apps leak in the payload of push notifications sent via FCM. Of these apps, \leakedAny{} leaked metadata, including user identifiers (\leakedUserIDs{} apps), sender or recipient names (\leakedNames{} apps), and phone numbers (\leakedPhones{} apps), while \leakedContents{} apps leaked the actual message content. Furthermore, none of the data we observed being leaked to FCM was specifically disclosed in those apps’ privacy disclosures. We also found several apps employing strategies to mitigate this privacy leakage to FCM, with varying levels of success. Of the strategies we identified, none appeared to be common, shared, or well-supported. We argue that this is fundamentally an economics problem: incentives need to be correctly aligned to motivate platforms and SDK providers to make their systems secure and private by default.
\end{abstract}

%% file: index.tex
\input{paper/1_introduction}

\input{paper/2_background}

\input{paper/3_related}
\input{paper/4_methods}
\input{paper/5_results}

\input{paper/6_discussion}
\input{paper/6a_disclosure}
\input{paper/7_limitations}

%% file: paper/1_introduction.tex
\section{Introduction}
\label{sec:intro}

\begin{quote}
{\it ``She speaks, yet she says nothing.''}

\hfill ---William Shakespeare, \textit{Romeo and Juliet}\\

\end{quote}

\input{assets/figures/example}



Modern economies rely on the specialization of labor~\cite{smith1776inquiry}. Software engineering is no different: modern software relies on myriad third-party components to fulfill tasks so that developers do not need to waste time rebuilding specific functions from scratch~\cite{Brooks1975}. This type of ``code reuse'' is a recommended practice and transcends many branches of engineering (e.g., car manufacturers do not manufacture every component that goes into their cars, instead relying on components from third-party suppliers). Software development kits (SDKs) facilitate code reuse during software development and offer many benefits for developers. They provide well-trodden paths: documented workflows for developers to follow so that these developers can consistently provide common functionality. Ultimately, SDKs reduce engineering costs when used responsibly.

Yet, recent research has demonstrated that many software privacy issues (i.e., the inappropriate disclosure of sensitive user information) are due to developers' misuse of third-party services~\cite{reyes2018won, alomar2022developers}. That is, privacy breaches often occur due to developers not correctly configuring SDKs, not reading SDK documentation, or SDKs behaving in undocumented ways, often unbeknownst to developers. This is especially concerning when the third-party SDK may transmit highly sensitive user data to third parties and the SDK is ubiquitous across many software supply chains.

Heightened public concerns around the monitoring of online communications have significantly influenced consumer behavior in the past decade. A 2014 PEW survey found that 70\% of Americans are concerned about government surveillance and 80\% about surveillance by corporations~\cite{Madden2014}. In response to these concerns, more and more consumers have begun using secure messaging apps to protect their communications based on the promises of privacy made by these apps. Hundreds of millions of users now use apps like Signal or Telegram, believing these apps to protect their privacy. These applications are entrusted with a vast array of confidential user data, from personal conversations to potentially-sensitive multimedia content, thereby placing a significant emphasis on their ability to make good on their promises of privacy and security.

The misuse of third-party SDKs within secure messaging apps may pose a heightened risk to users because those SDKs may leak sensitive information to third parties. In particular, app developers use third-party SDKs to implement \textit{push notifications}, which display important information to the user, including messages from other app users (Figure \ref{fig:push}). Because push notification SDKs are generally provided by third parties (as opposed to app developers), incorrect usage may leak sensitive information to those third parties. For example, an app that provides ``end-to-end'' encrypted messaging may not actually provide end-to-end encryption if message payloads are not encrypted before being sent to third-party push notification APIs. To make matters worse, misuse of these SDKs may also contribute to the misrepresentation of security and privacy assurances to consumers as articulated in various disclosures, including privacy policies, terms of service, and marketing materials.

The combined risk of sensitive information leakage and misrepresentation of privacy promises creates serious ramifications for users of secure messaging platforms. Oppressive regimes or other adversaries may use court orders to compel companies involved in the delivery infrastructure of push notifications to reveal the contents of communications sent and received by human-rights workers, political dissidents, journalists, etc. Worse, when this does happen, both the developers of the apps and the users who are endangered are unlikely to be aware that their communications are being intercepted. This threat model is not just theoretical. Crucially, since we performed our analysis, U.S. Senator Ron Wyden published an open letter that confirms that government agencies do, in fact, collect user information by demanding push notification records from Google and other push notification providers through the use of legal processes~\cite{wydenLetter}. Our work is highly prescient, as it provides new insights into an emergent threat model. 

To study the extent to which the delivery infrastructure may access sensitive user information, we examined the use of Google's \textit{Firebase Cloud Messaging} (FCM) to deliver push notifications to \textit{secure messaging apps} on Android devices. Google provides FCM as a free service, and therefore, it is one of the most commonly used third-party SDKs to deliver Android push notifications. Moreover, the majority of other push services, including OneSignal~\cite{oneSignalStatement}, Pusher~\cite{pusherStatement}, and AirShip~\cite{airshipStatement} internally rely on Google's FCM to deliver notifications to Android devices, making the usage of FCM practically unavoidable for developers who wish to provide push notification support in their Android apps.  (On Apple's iOS, third-party push notification APIs are similarly built on top of Apple's push notification service~\cite{oneSignalPush}.) We focus on secure messaging apps because these apps (1) market their abilities to keep message data ``private'' or ``secure'' and (2) make heavy use of push notifications to notify users of incoming messages and their contents (and therefore, when not implemented correctly, may run the risk of leaking message contents and metadata to the push notification service). 

Prior work has investigated the potential security risks that push notifications may pose, including by push notification-based malware~\cite{hyun2018design, li2014mayhem} and botnets~\cite{lee2014punobot, hyun2018design}. To our knowledge, no work has focused on the privacy risks of push notification services used by secure messaging apps. Therefore, we performed a study to examine whether the push notification records potentially stored without end-to-end encryption by the delivery infrastructure may misrepresent or compromise the privacy protections of secure messaging and expose users to legal risks. Thus, we posed the following research questions:

\begin{itemize}
    \item \textbf{RQ1:} What personal data do secure messaging apps for Android send via Google’s Firebase Cloud Message (FCM)?
    \item \textbf{RQ2:} What mitigation strategies do app developers use to protect personal information from being disclosed to Google’s FCM?
    \item \textbf{RQ3:} Do the observed data-sharing behaviors align with the privacy assurances apps make in their public disclosures?
\end{itemize}

To answer these questions, we performed static and dynamic analysis on a corpus of \apps{} secure messaging apps. We used dynamic analysis to understand what data these apps sent over the network. When we found that apps displayed data in push notifications, but did not obviously send that data over the network, we used static analysis to understand what mitigation strategies they used to achieve this effect. In contrast, when segments of data displayed in the app \emph{were} verbatim in push notifications, we further examined these messages to assess whether sensitive data was available in plaintext to the delivery infrastructure. Finally, we analyzed apps' privacy policies and other disclosures to identify the privacy claims that apps made to users. By comparing observed behavior from our app analysis to disclosed behavior, we identify undisclosed sharing and potentially-misleading data practices: data that apps imply that they will not disclose, but---intentionally or not---do disclose to the delivery infrastructure through the use of push notifications.

We found that more than half of the apps in our corpus leak \textit{some} personal information to Google via FCM\@. Furthermore, none of the data we observed being leaked to FCM was specifically disclosed in those apps’ privacy disclosures. We also found several apps employing strategies to mitigate this privacy leakage to FCM, with varying levels of success. Of those identified strategies, none appeared to be common, shared, or well-supported. While app developers are ultimately responsible for the behavior of their apps, they are often ill-equipped to evaluate their apps' privacy and security properties in practice. Given that the problems that we observe are pervasive across app developers and stem from the use of third-party components that can be easily used insecurely, we conclude that SDK providers are best positioned to fix these types of issues through both better guidance and privacy-preserving designs and defaults.


In this paper, we contribute the following:
\begin{itemize}
  \item We demonstrate the widespread sharing of personal information, perhaps inadvertently, with Google through developers' use of push notifications.
  \item We highlight systemic mismatches between privacy disclosures and observed behaviors in delivering push notifications via FCM.
  \item We discuss developers' negligence in deploying software that they do not understand and the responsibility that SDK and platform providers share in creating infrastructures that are private/secure by default.
\end{itemize}

%% file: assets/figures/example.tex
\begin{figure}[t]
  \centering
  \includegraphics[width=0.8\columnwidth]{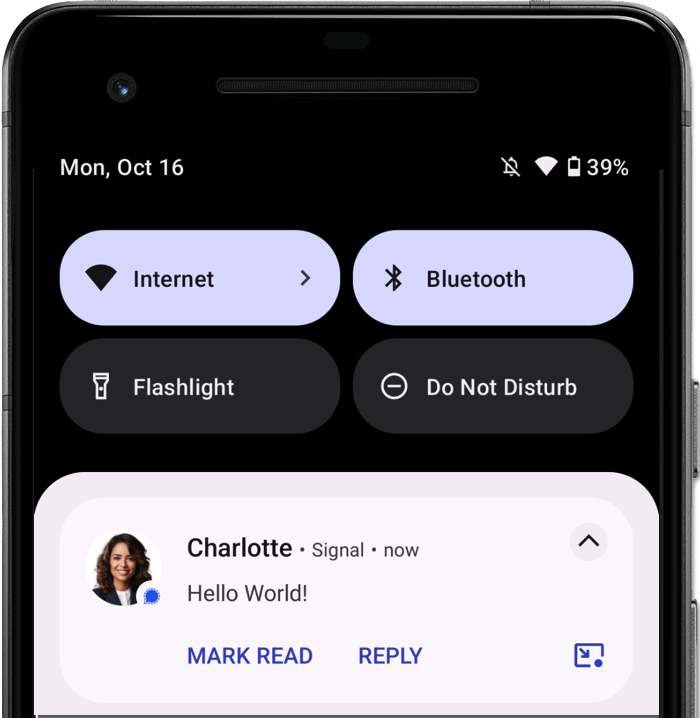}
  \caption{An illustration of an Android push notification.}
  \label{fig:push}
\end{figure}

%% file: paper/2_background.tex
\section{Background}
\label{sec:background}

We provide an overview of push notification services (PNS), specifically Google's Firebase Cloud Messaging (FCM). We describe the threat model we consider in this paper and our overall motivation.

\subsection{Mobile Push Notifications}
A push notification is a short message that appears as a pop-up on the desktop browser, mobile lock screen, or in a mobile device's notification center (Figure~\ref{fig:push}). Push notifications are typically opt-in\footnote{Android and iOS require user permission before an app can display notifications.} alerts that display text and rich media, like images or buttons, which enable a user to take a specific action in a timely fashion, even when the app in question is in the background. Applications often use push notifications as a marketing or communication channel, but they can also be used as a security mechanism (e.g., as part of a multi-factor authentication ceremony).

There is a difference between push messages and notifications. ``Push'' is the technology for sending messages from the server-side component of the app (the ``app server'') to its client side (the ``client app''), even when the user is not actively using the app. Notifications refer to the process of displaying timely information to the user by the app's user interface (UI)~\cite{pushOverview}. In the context of mobile apps, the application server can send a push message without displaying a notification (i.e., a silent push); an app can also display a notification based on an in-app event without receiving any push messages. For simplicity's sake, we use the term ``push notifications'' in this paper regardless of whether an actual notification is displayed to the end user (i.e., we refer to messages flowing through a cloud messaging server to a user's device, whereupon the device's operating system routes the messages to the appropriate app). 

\begin{figure}[t]
    \centering
    \frame{
        \includegraphics[width=0.55\columnwidth]{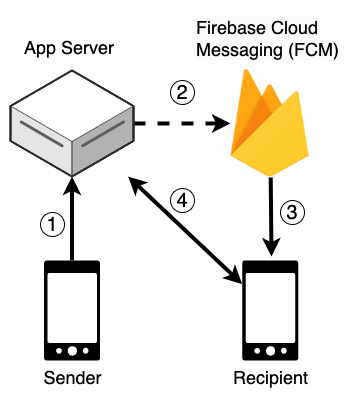}
    }
    \caption{Flow chart of FCM's push notification infrastructure for messaging apps, highlighting the actors involved and the interactions between them: an event occurs that triggers a push notification, e.g., a message from a sender (1) prompts the app server to create and send the message to FCM (2), which then forwards it to the recipient's Android device (3). If needed, the receiving app running on that device may also query additional information from the app server (4).}
    \label{fig:push-notification-diagram}
\end{figure}

Although app developers could, in theory, implement their own push notification service, this is usually impractical as it requires the app to continually run as a background service, thereby reducing battery life. Instead, most mobile app developers rely on \textit{operating system push notification services} (OSPNSs), including Firebase Cloud Messaging (FCM) for Android or Apple Push Notification Service (APNS) for iOS devices~\cite{AppleNotifications}. FCM and other PNSs facilitate push notifications via an SDK the developer adds to their application. When a user launches the app for the first time, the SDK registers the device with the PNS by generating a \textit{push token} (also known as a \textit{registration token}), which serves as a pseudonymous identifier that tells the push service where to forward the messages. The SDK returns the push token to the client app, which should then be sent and stored in a database on the app server. When the app wants to send a push notification, it looks up the appropriate push token and sends it alongside the message to the PNS, which then forwards the message to the correct device~\cite{wickham2018push}. The push token is tied to the app instance, and therefore, the developer should periodically refresh it, e.g., if the user deletes and reinstalls the app.  

In sum, there are three main actors involved in the process of sending push notifications using FCM (see also Figure~\ref{fig:push-notification-diagram}):

\begin{description}

\item[App Server] sends event-specific messages to FCM (2). For instance, in the context of a messaging app, a sender device may send a message to the app server (1), which then sends a push notification request to FCM (2).
\item[Firebase Cloud Messaging (FCM)] is a cloud-based OSPNS that forwards push messages to the appropriate user device using the stored registration token(3), even if the client app is offline or in the background. It also exposes an API to the developer to enable push messaging in their applications. 
\item[Android Device] runs the OS and the client app. Android uses a system component that is part of Google Play Services to receive push messages sent by FCM, which it then passes to the appropriate app. Optionally, the client app can also query additional information from the app server (4) in response to a received push notification.
\end{description}

The SDKs distributed by FCM and other PNSs not only streamline app development by reducing the amount of code that needs to be written, but in many cases, their use is necessary for performance and efficiency reasons~\cite{telegramPersistentPush}. Developers would also need to request the Android permission for unrestricted battery usage, something a user might not necessarily grant. As such, mobile platform owners only provide official support for their managed OSPNSs: Google for FCM and Apple for ASPNS.\footnote{We studied Android because the operating system is open source, allowing us to more easily build instrumentation to monitor app execution.}

\subsection{FCM Alternatives}
\label{sec:alternatives}
Given the utility of push notifications, companies have started offering push notification services that compete with Google's FCM. These third-party PNS providers, such as Airship, Pushwoosh, and Taplytics, may offer advantages over FCM, including more features or usable APIs. While it may seem that developers using third-party PNSs can potentially avoid the security and privacy pitfalls of FCM, Lou et al.\ demonstrated that third-party push providers rely on FCM to deliver messages to Android devices with Google Play Services~\cite{lou2023devils}. The authors identified the dual-platform structure of push notifications. The first service (``host notification platform'') abstracts push messaging by providing an API that interfaces with the second service (``transit notification platform''), which provides a stable system-level communication channel to deliver push notifications to user devices. While both FCM and third-party PNSs offer developer-facing APIs for managing push notifications (i.e., the host notification platform), only FCM fulfills the role of the transit notification platform and delivers messages internally to Android devices with Google Play Services. 

Furthermore, we found statements by several popular third-party PNSs, such as OneSignal~\cite{oneSignalStatement}, Pusher~\cite{pusherStatement}, and AirShip~\cite{airshipStatement} that mention their dependence on FCM for sending push notifications to Android devices. For instance, OneSignal states in a blog post that ``Google mandates that Android apps distributed through Google Play leverage a single, shared connection provided by FCM'' and ``OneSignal itself uses the FCM API internally to send messages to Android devices''~\cite{oneSignalStatement}. Therefore, these third-party PNSs expose users to risks associated with FCM push notifications while potentially introducing their own problematic data collection practices. For instance, Reuters has previously reported that Pushwoosh---a third-party PNS---misrepresented itself as based in the U.S. despite actually being headquartered in Russia~\cite{reutersPushwoosh}. Although Pushwoosh denied the claims~\cite{pushwooshStatement}, the revelation still led the U.S. Army and Centers for Disease Control and Prevention (CDC) to stop using apps containing the Pushwoosh SDK.

Android devices without preinstalled Google Play Services either do not properly support push notifications or use an alternative platform. Most notably, Android devices sold in China do not include Google Play Services, but use another preinstalled service provided by the phone manufacturer, such as Huawei Mobile Services (HMS), to handle push notifications. There are other Android variants outside of China that do not come with Google Play Services preinstalled, such as FireOS, which runs on Amazon devices and uses Amazon Device Messaging (ADM) instead of FCM. These variants constitute a small share of the global Android market~\cite{gizChinaStatement} and are outside the scope of our analysis.

Other alternatives, such as UnifiedPush~\cite{unifiedPush} or Samsung Push Service~\cite{samsungPush}, rely on apps to receive push notifications in place of Google Play Services. However, we argue that such solutions do not represent equivalent alternatives, as they require users to install an additional app and developers may still use FCM as the push service, unbeknownst to app users. Thus, we specifically focus on data shared with Google's FCM, regardless of the specific third-party service running on top of it. (That is, our instrumentation is agnostic as to whether it captured messages sent natively using FCM or another third-party API built upon it.)



\subsection{Threat Model}
FCM acts as an intermediary between the server-side and client-side applications and uses push tokens to identify the device where push notifications should be forwarded. While efficient, this architecture poses three significant privacy risks to users~\cite{wuyts2020linddun, enisaSharing}:

\begin{description}
    \item[Disclosure.] The contents of a push notification and its metadata may be disclosed to unauthorized entities.
    \item[Linking.] Push tokens may be linked or attributed to specific users or behaviors. 
    \item[Identification.] Individuals may become identified based on the information linked to their device's push tokens. 
\end{description}

The primary threat model that we consider is the use of legal processes to request FCM push tokens linked to a targeted device and stored by the app developer. In the context of secure messaging apps, knowing the pseudonym (i.e., username) of the targeted user may suffice. Even if the app developer does not collect other identifying personal information, they must still store registration tokens to route the push notifications to the user’s device through FCM servers. After obtaining the push tokens from the app publisher, law enforcement can request that Google provide all information linked to the given push token, which may include the contents and metadata of the associated push notifications. Combining these pieces of personal information increases the risk of identification.

This threat model is not theoretical. In December 2023, U.S. Senator Ron Wyden published an open letter confirming that government agencies collect user information by demanding push notification records from Google and Apple through legal processes~\cite{wydenLetter}. Since then, journalists found more than 130 search warrants and court orders going back to 2019 (e.g., ~\cite{pushWarrant1, pushWarrant2, pushWarrant3}) in which investigators had demanded that tech companies, notably Wickr and TeleGuard---both advertised as end-to-end encrypted secure messaging apps---turn over push tokens associated with accounts of specific users. In the case of TeleGuard, an FBI agent then asked Google to hand over all information connected to the push token, which Google responded to with account names and IP addresses associated with those accounts~\cite{wapoPushThreats}. Furthermore, Apple disclosed in its transparency report for the second half of 2022 that it received 70 requests worldwide seeking identifying information about Apple Accounts (formerly known as Apple IDs) associated with 794 push tokens and provided data in response to 54 (77\%) requests. Google does not specifically break out government requests for push notification records and, instead, reports these requests in aggregate with other account data requests~\cite{appleTransparencyReport}.  

We hypothesize that many Android app developers transmit sensitive information via established third-party push notification channels and do not realize that they are not properly securing it. In a departure from ``privacy-by-design'' principles~\cite{cavoukian2009privacy}, the official Google Android Developers Blog recommends~\cite{androidDevBlog} that developers using Google's service ``send as much data as possible in the [push notification] payload'' and fetch the remainder of the data from the app server if needed. In the next paragraph of the blog, developers are advised that they ``can also encrypt FCM messages end-to-end using libraries like Capillary,'' thereby indicating that FCM does not encrypt payload data by default (i.e., developers need to rely on additional libraries). There is no other mention of end-to-end encryption in the blog. Thus, questions remain as to whether developers follow this optional guidance.

Google's FCM developer documentation~\cite{fcmMessage} states that ``depending on your needs, you may decide to add end-to-end encryption to data messages'' and ``FCM does not provide an end-to-end solution.'' No further guidance is given on what information is appropriate to send. In contrast, Apple's documentation for sending notifications~\cite{AppleNotifications} instructs developers not to include ``customer information or any sensitive data in a notification's payload'' and, if they must, ``encrypt it before adding it to the payload.'' Even if the majority of data sent using push notification channels is not personal, there are examples in which it might be, such as some user-generated content in instant messaging apps or sensitive information sent by a banking or a health-tracking app. In these cases, app vendors may be held liable for failing to safeguard or minimize the amount of personal information sent via push notification servers and for failing to disclose this practice in their privacy notices. 

\input{assets/figures/fcmblog}

Given FCM's role as an intermediary, we posed the question: do apps leak user information through push notifications to the delivery infrastructure? We investigated this question by performing both mobile app analysis and analysis of privacy disclosures.

%% file: assets/figures/fcmblog.tex
\begin{figure}[t]
  \centering
  \fbox{\includegraphics[width=0.98\columnwidth]{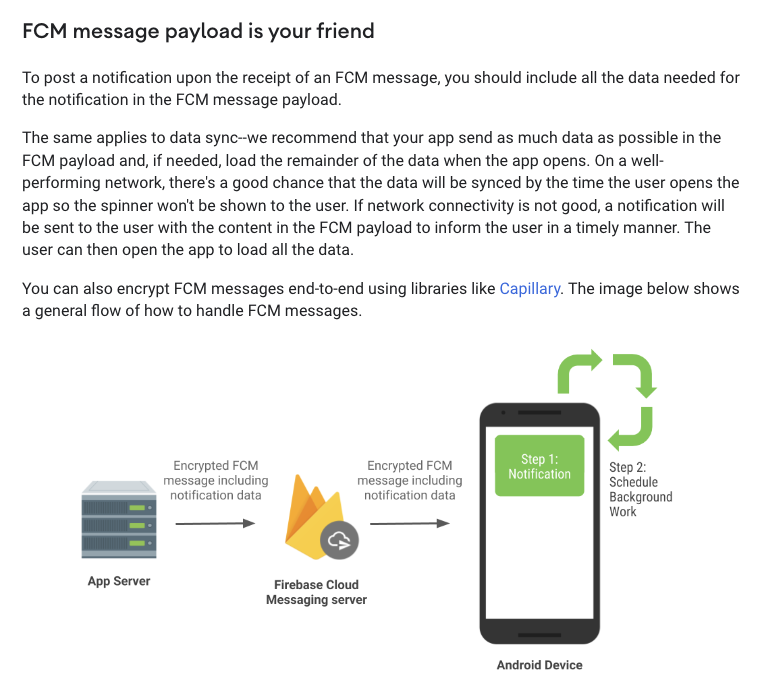}}
  \caption{Google's guidance to send as much data as possible via FCM payloads, noting that end-to-end encryption can optionally be used via additional libraries~\cite{androidDevBlog}. It is unclear whether the data flows labeled ``encrypted'' refer to this option or the fact that the transmissions use TLS.}
  \label{fig:fcmblog}
\end{figure}

%% file: paper/3_related.tex
\section{Related Work}
\label{sec:related}

In this section, we provide an overview of related work on the privacy and security risks of push notifications, mobile app analysis, and analysis of privacy-relevant disclosures.

\subsection{Risks of Push Notifications}
Prior research has demonstrated how attackers can exploit mobile push notifications to spam users with advertisements~\cite{liu2019dapanda}, launch phishing attacks~\cite{xu2012abusing}, and even issue commands to botnets~\cite{ahmadi2016detecting, lee2014punobot, hyun2018design}. Other studies have revealed additional security issues with PNSs that can result in the loss of confidentiality (i.e., user messages get exposed to unauthorized parties) and integrity (i.e., users receive malicious messages from unauthorized parties)~\cite{chen2015perplexed}. By assuming that the victim installs a malicious app, prior work has demonstrated how attackers can abuse platform-provided OSPNSs, including Google’s FCM (formerly known as Google Cloud Messaging or GCM, and Cloud to Device Messaging or C2DM prior to that), to steal sensitive messages or even remotely control the victim’s device~\cite{li2014mayhem}. Warren et al.\ described ``security'' as a key nonfunctional requirement for implementing push notification mechanisms and identified the push-to-sync strategy back in 2014 (which they called ``poke-and-pull'') as a viable protection strategy for protecting user data from PNSs~\cite{warren2014push}. 

As described previously (\S~\ref{sec:alternatives}), push notification architecture can be separated into the host platform that provides the push API and the transit platform that actually delivers the push notification internally. Several studies looked at the security issues of third-party PNS SDKs while excluding system-level transit platforms, such as FCM from Google. One study analyzed 30 different third-party PNS SDKs embedded in 35,173 Android apps and found that 17 SDKs contain vulnerabilities to the confidentiality and integrity of push messages, which an attacker can exploit by running a malicious app on the victim’s device~\cite{chen2015perplexed}. Similarly, Lou et al.\ performed a security and privacy analysis of the twelve most popular PNSs and compared their behavior in 31,049 apps against information practices disclosed in the privacy policies of those PNSs~\cite{lou2023devils}. They found that out of twelve third-party PNSs, six PNSs collect in-app user behavior and nine collect location information, often without awareness or consent of app users. As the authors focused only on the host platforms, their analysis excluded FCM (and other transit platforms) on the basis of them being a ``trustful service provider.'' We complement this work by focusing instead on the privacy risks of transit notification platforms, in particular, FCM from Google. 

In recent years, researchers have analyzed PNSs from the perspective of privacy protection goals that complement the classic ``CIA triad'' (confidentiality, integrity, and availability), such as unlinkability, transparency, and intervenability~\cite{hansen2015protection}. One study, for instance, considered an adversary with the capability to silently sniff packets directed to or from the victim and actively trigger push notification messages to the target’s personal device~\cite{loreti2018push}. The authors demonstrated that under these assumptions, an adversary on the same network can deidentify the victim even if they use an online pseudonym. We complement these studies by focusing on FCM privacy risks in the context of secure messaging apps.

\subsection{Mobile App Analysis}
Numerous studies have also investigated the security and privacy ramifications of mobile apps (e.g.,~\cite{Felt2012, kelley2013privacy, Thompson2013, Tan2014}). Most current methods for evaluating mobile app actions depend on static analysis~\cite{kim2012scandal, Gibler2012, Gordon2015, zimmeck2017automated}, which examines the app's source code without executing it. However, this technique is limited as it can only identify the potential behaviors of a program, not if and to what degree the program exhibits them. For instance, it is generally infeasible to predict the full set of execution branches that a program will take. Alternative methods, such as taint tracking~\cite{Enck2010}, which tracks the flow of data as it propagates through the application, come with their own challenges, including affecting app stability~\cite{Cavallaro2008}.

A newer approach involves adding instrumentation to the Android operating system to monitor apps' access to personal information at runtime~\cite{wijesekera2015android, wijesekera2017feasibility, Tsai2017, wijesekera2018contextualizing}. This allows researchers to investigate different app behaviors, including app-associated network traffic. Prior solutions to monitoring mobile app transmissions generally involve using proxy software (e.g., Charles Proxy,\footnote{\url{https://www.charlesproxy.com/}} mitmproxy,\footnote{\url{https://mitmproxy.org/}} etc.) and suffer from serious shortcomings. First, they route all the device traffic through the proxy, without automatically attributing traffic to a specific app running on the device. While some traffic may contain clues (e.g., content and headers that may identify apps, e.g., HTTP \texttt{User-Agent} headers), other traffic does not, and attributing traffic to the app is a laborious and uncertain process~\cite{razaghpanah2017studying}. Second, proxies often cannot automatically decode various obfuscations, including TLS with certificate pinning. Instead, by capturing traffic from the monitored device's OS, these issues are eliminated. This approach can bypass certificate pinning, extract decryption keys from memory, and map individual sockets to process names, thereby offering precise attribution to specific apps.

\subsection{Analysis of Privacy Disclosures}
Prior research has focused on understanding apps' and websites' privacy practices by analyzing disclosures made in privacy policies~\cite{harkous2018polisis, andow2020actions, wang2018guileak, zimmeck2019maps, zimmeck2017automated}. Some proposed systems, such as \textsc{policheck}~\cite{andow2020actions}, \textsc{maps}~\cite{zimmeck2019maps} and \textsc{hpdroid}~\cite{fan2020empirical}, which automate the process of comparing disclosures made in privacy policies about how user data is used, collected, or shared with personal data transmissions observed as a result of performing technical analyses~\cite{wang2018guileak, andow2020actions,zimmeck2019maps, zimmeck2017automated, slavin2016toward}. The literature also proposed systems, such as Polisis~\cite{harkous2018polisis}, PI-Extract~\cite{bui2021automated} and PrivacyFlash~\cite{zimmeck2021privacyflash}, which made it possible to transform privacy policies into formats that are more understandable to users or auto-generate policies that reflect actual app behaviors. Linden et al.~\cite{linden2018privacy} found that disclosures made in privacy policies improved as a result of GDPR enforcement, but that more improvements would have to be made before they can be considered usable and transparent to users. Other recent studies have also examined the accuracy of disclosures made in privacy policies~\cite{andow2019policylint, okoyomon2019ridiculousness, wang2018guileak, samarin2023lessons}. 


Additionally, Google's Play Store requires developers to provide privacy labels~\cite{googlePrivacyLabels}. Privacy labels communicate information practices to users in a visually succinct way. For example, apps may list the data types (e.g., names, phone numbers, identifiers) collected and shared with third parties. As with privacy policies, these privacy labels are required by the Google Play Store's terms of service to be thorough and complete~\cite{googlePrivacyLabels}. However, Google states in their guidelines that ``transferring user data to a `service provider'{}'' should not be disclosed as data sharing in the app's privacy labels~\cite{googlePrivacyLabels}, limiting their scope and potential utility. Other studies have also demonstrated the inconsistencies between privacy labels and privacy policies~\cite{stoppersee}, privacy labels in the Google Play Store and Apple App Store for the same apps~\cite{rodriguez2023comparing}, and practices disclosed in privacy labels and behaviors observed among iOS apps~\cite{koch2022keeping, xiao2022lalaine}.





%% file: paper/4_methods.tex
\section{Methods}
\label{sec:methods}


Our primary research question concerns how secure messaging apps' usage of FCM impacts user privacy. To answer this question, we identified a set of apps from the Google Play Store and compared the claims made in their privacy disclosure documents with our static and dynamic analysis of those same apps.

The diagram in Figure~\ref{fig:push-notification-diagram} outlines the main actors and communications involved in push notification usage in secure messaging apps. The messaging app is installed on the phone/device of the sender and the recipient. First, the sender composes their message, and some content gets sent over the network to the app's server (1). Then, the server uses the FCM API to construct the push notification with the required payload. The FCM API sends the notification to Google's FCM server (2), which then forwards it to the recipient device (3) using a long-lived TCP connection initiated by Google Play Services. Finally, the data is parsed and packed into an intent that is then broadcast to the app, which displays the message in the form of a notification. Inadvertent data leakage to Google occurs when the server places user information as plaintext in the push notification payload. Crucially, users and developers are likely unaware that Google may receive and, sometimes, retain\footnote{E.g., FCM servers retain messages by default when the recipient device is offline.} message contents and other metadata associated with the push notification.

As highlighted in \S~\ref{sec:related}, numerous prior works evaluate the security and privacy of end-to-end (e2e) encryption and its implementation in secure messaging apps, including many of the ones in our corpus. However, our work is explicitly \textbf{not} investigating these claims of e2e encryption. Therefore, we are not interested in recording the traffic sent over a network connection. Rather, our interest is in determining whether implementing push notification functionality in a given app leaks personal message content to parties \textit{other than the app developer}, specifically to Google via FCM. Therefore, we are primarily interested in what data the app's server sends to FCM via network connection. However, because we are out-of-band from this network connection, the best alternative is to record the inbound/outbound traffic on the recipient's device to infer which data may have been sent from the server to FCM. If the sender's plaintext message content is present in the push notification sent to the recipient's device from FCM, then it is clear that the app server did leak the user's message content to FCM. However, if the push notification sent to the recipient's device does not contain the sender's plaintext message, then it may be likely that the app server did \textbf{not} leak the user's message content to FCM.\footnote{If the app server has access to the sender's plaintext message, then it is always possible that it is leaked to third-parties in ways that are not externally detectable, since traffic between the app server and these third parties is not observable.} For apps that fall into this category, we additionally want to understand the techniques they leverage to avoid leaking user message content and metadata to FCM.

\subsection{App Selection}
We selected messaging apps that made claims about the privacy of users' messages (herein, ``secure messaging apps''). For example, Telegram's homepage promotes its app as ``private'' and states that ``Telegram messages are heavily encrypted''~\cite{telegramMainPage}. Similarly, Signal's homepage encourages people to ``speak freely'' because the Signal app has a ``focus on privacy''~\cite{signalMainPage}. Signal publicly writes about what data their app collects and the fact that---in response to a legal subpoena requesting a range of user information---Signal is only able to provide ``timestamps for when each account was created and the date that each account last connected to the Signal service''~\cite{signalGrandJury}. WhatsApp also explicitly markets the privacy benefits of their app and states, ``[y]our privacy is our priority. With end-to-end encryption, you can be sure that your personal messages stay between you and who you send them to''~\cite{whatsappBillboard, whatsappFunApple}. Because secure messaging apps make these claims about the privacy of users' messages, many users utilize these apps in sensitive contexts. For example, Telegram, Signal, and WhatsApp, three of the apps we analyzed, are frequently used by protesters worldwide~\cite{slobozhan2023differentiable, urman2021analyzing}. The apps in our data set, a subset of all secure messaging apps, are widely used and encompass over 2.8 billion users and 6.1 billion installs.

\mysubsubsection{Material Representations}
The selection of messaging apps based on their privacy claims is not only a prudent approach for users prioritizing the confidentiality of their communications, but also a legally-grounded strategy, reflecting the enforceable nature of such assertions. When companies publicly assert their services' privacy and security features, these claims become material representations that can significantly influence consumer choices. Importantly, material misrepresentations are actionable under consumer protection laws. For instance, under the FTC Act\footnote{15 U.S.C. \S45.} (and various state consumer protection laws), businesses in the U.S. are prohibited from materially misrepresenting their practices to consumers. The Federal Trade Commission (FTC) and state attorneys general actively monitor and pursue companies that fail to uphold their privacy promises (regardless of whether they are made in privacy policies~\cite{FTCFlo2021} or marketing materials~\cite{FTCAvast2024}). This enforcement protects consumers and reinforces the message that privacy and security assertions are material representations that have legal consequences and can affect consumer choices.

One such notable case is that of Zoom, in which the company faced a regulatory enforcement action for erroneously claiming to offer end-to-end encryption in its marketing materials, a feature it did not fully provide at the time~\cite{ftcStatement}. This incident underscores the seriousness with which authorities treat misrepresentations in the digital privacy domain, highlighting the risks companies face when they do not accurately describe their data protection measures. Thus, evaluating messaging apps based on their stated privacy features is not only a measure of their utility in sensitive contexts, but also an assessment of their compliance with legal standards for truthfulness in advertising, ensuring that users can rely on the integrity of these claims.


\mysubsubsection{Selection Procedure}
We aimed to create a corpus of secure messaging apps that made privacy claims to users, such that it included widely-used apps and was of a tractable size to perform our analyses. To create this corpus, we first had to identify a set of the most popular secure messaging apps in the Google Play Store. We focused on apps in the \texttt{Communication} category in the Google Play Store, which included a broad range of messaging apps, including email clients, mobile browsers, and SMS apps. Within this category, we used open-source tooling\footnote{\url{https://github.com/facundoolano/google-play-scraper}} to identify apps whose descriptions included one or more keywords related to online messaging\footnote{``messaging,'' ``chat,'' ``internet,'' ``friend,'' and ``in touch.''} and explicitly excluded keywords related to non-messaging apps.\footnote{``SMS,'' ``browser,'' ``VPN,'' ``recover,'' and ``voicemail.''}

To establish this list of keywords, we manually reviewed the descriptions of apps in the \texttt{Communication} category and iteratively added keywords to our inclusion and exclusion lists until we manually determined that the resulting set of apps included secure messaging apps that do not fall back onto SMS. Then, we excluded any app whose description did not include the terms ``privacy'' or ``security.'' Finally, we only selected apps with more than a million installations. This penultimate set contained \selectedIMApps{} apps. We decided not to analyze Google Messages because it is owned by Google and, therefore, there is no notion of third-party leakage in that app; Google runs the infrastructure that provides the push notifications. We also excluded Leo Messenger, which appeared to aggregate other messaging apps and did not have messaging functionality in its own right, as well as Gap Messenger, for which we were unable to register. Therefore, the final set contained \apps{} apps.

\subsection{App Analysis}
We performed dynamic and static analysis on each secure messaging app in our data set to learn how the usage of FCM impacted user privacy. Specifically, did the app na\"ively leverage the default FCM behavior and include plaintext user content? Or, did the app use specific techniques to protect the privacy of user messages above and beyond what FCM offers by default? (For example, by integrating the Capillary library~\cite{capillary} mentioned in Google's blog.)

\mysubsubsection{Data Types}
In our analysis, we searched for specific data types that we expected to appear in the content of push notifications. To compile the list of these data types, we started with the data types defined and used by Google's privacy labels~\cite{googlePrivacyLabels}, which also enabled us to compare observed practices with the privacy labels declared by each app's developer. As we present in Section~\ref{sec:results}, we found evidence of the following data types being leaked to Google: (1) \textit{Device or other IDs}, (2) 
\textit{User IDs}, (3) \textit{Name}, (4) \textit{Phone Number}, and (5) \textit{Message Contents}. Unlike (1) to (4), the contents of communications are afforded additional protections in many jurisdictions due to their sensitive nature.\footnote{E.g., Title I of the Electronic Communications Privacy Act of 1986 (ECPA)~\cite{ecpa}.} We present additional information about these data types in Appendix~\ref{sec:datatypes}.



We performed our analysis in early 2023 with an instrumented version of Android 12, at a time when the majority of users (more than 85\%) had Android version 12 or below installed on their phones~\cite{androidMarketShare}. Using a Pixel 3a phone, we installed each app from Google Play Store and saved its Android package (APK) files and privacy disclosures. We also created test accounts where necessary. We then used dynamic analysis to identify what personal information got leaked to FCM and static analysis to understand what strategies apps used to protect user privacy.

\mysubsubsection{Data Leakages}
We used dynamic analysis to record the contents of a push notification after our device received it from the FCM server. We instrumented the \texttt{keySet()} method of the standard \texttt{BaseBundle} class~\cite{baseBundle}, which gets called by the FCM SDK, and logged the contents of the \texttt{Bundle} only if it contained the default keys in a push notification, such as ``google.message\_id.'' Additionally, we used Frida~\cite{Frida} to instrument the \texttt{handleIntent} method of \texttt{FirebaseMessagingService}~\cite{firebaseMessagingService}, which listens and receives FCM push notifications as broadcasts from Google Play Services. This method then delivers push notification contents to app-specific callback methods (e.g., \texttt{onMessageReceived}), which allow the app to handle and display push messages as notifications to users.

The main goal was to trigger a push notification so that the resulting payload sent from Google's FCM server to our test device could be recorded (connection 3 in Figure~\ref{fig:push-notification-diagram}).  We installed each app on two devices and triggered push notifications by sending messages from one device to another. On the recipient's Pixel 3a device, we recorded the push notification contents as they were received by the app using the instrumented methods. 

\mysubsubsection{Privacy Strategy}

The push notifications that we observed fell into one of the following three categories:

\begin{enumerate}
    \item \textbf{No Protection.} The FCM push notification contained all of the information (i.e., username and message contents) that the app uses to display the notification.
    \item \textbf{Some Protection.} The FCM push notification contained some personal information but, notably, did not include the displayed message contents in plaintext.
    \item \textbf{Full Protection.} The FCM push notification did not contain any personal information, and any additional fields were empty or always contained unique values (i.e., not corresponding to any persistent identifiers). 
\end{enumerate}

For the first case, we simply assumed that the app does not use any privacy protection strategies. For the latter two cases, determining the strategy was often straightforward. For instance, Skype (in secret chat) included \texttt{EndToEndEncryption} as the value for the \texttt{messagetype} key, while Session included the \texttt{ENCRYPTED\_DATA} key with a value corresponding to an encoded message. Signal, on the other hand, received FCM push notifications that only contain the empty field \texttt{notification} without any other content. 


To validate the identified strategies, we performed static analysis. We first decompiled the APKs for each closed-source app using the \texttt{jadx}\footnote{\url{https://github.com/skylot/jadx}} Dex to Java decompiler. Analyzing obfuscated code was often complex. We searched for \texttt{FirebaseMessagingService} to find services that extend it. We then examined the code of these services to see how they implement the \texttt{onMessageReceived} method, which gets invoked by the FCM SDK whenever the app running on the client device receives a push notification. Crucially, the SDK also passes a hash table of type \texttt{RemoteObject} containing information necessary to display the notification to the user and, optionally, a data payload to perform any custom functions triggered by the receipt of a notification. 

We tried to determine whether the push notifications contain sensitive content by observing the strings defined in code and used in the names of the keys or in print statements. We then traced the message and any variables assigned to the sensitive content until we reached the code for displaying the notification to the user. Appendix~\ref{sec:workflow} includes the questions we used to analyze the source code of apps in our data set.

\input{assets/tables/table_broken_promises}

\subsection{Privacy Disclosure Analysis}
The final phase of our analysis involved comparing the claims that app developers made in their privacy disclosures to the ground truth that we observed from our dynamic and static analysis. Therefore, we focused on the \leakedAny{} app developers that we observed including personal information in the push notifications sent via Google's FCM (\S~\ref{sec:results}). We wanted to determine whether a user could reasonably conclude that the app guarantees the security and privacy of their personal information based on the information presented by the app vendor in their Play Store description, official website, marketing and promotional materials, and other documentation. Moreover, we wanted to understand whether developers disclose the sharing of personal information for the purposes of providing push notifications in their privacy policies.

To achieve this, several researchers from our team first located statements by app vendors that talk about the security and privacy of messages. We also determined whether the apps (that we observed sharing personal information with Google) claimed to support end-to-end encryption by default, potentially misleading the users about the privacy of their messages or their metadata. Finally, we read each privacy policy to determine whether they stated that the particular types of personal information we observed might be shared with service providers for the purpose of app functionality. If it did, we further recorded whether the privacy policy listed the specific service providers or the specific types of data shared for the purpose of app functionality, which we compared against the results of our app analysis. By cross-referencing the different sources of information about an app's privacy practices, we aimed to build a holistic picture of how each developer frames the privacy risks associated with use of their app. We saved static copies of each privacy disclosure and the privacy policies using the Internet Archive's Wayback Machine~\cite{waybackMachine}.

\subsection{Ethical Research}
Our work involves reverse-engineering the client apps of popular Android secure instant messengers in order to glean the types of information being leaked to Google's FCM servers in push notifications. We performed our analysis by running each app on our test devices, with test accounts, on a segmented and private network, and observing both the network traffic that resulted and, when that network traffic did not reveal personal information, the static code. We were only interested in observing the leakage of personal information pertaining to our test devices; we did not interact with other app users nor did we make any attempts to obtain personal information of other users. Our study did not involve human subjects, nor did it involve unauthorized access to protected systems.

As we discuss in Section~\ref{sec:results}, we found inconsistencies between the observed app behavior and promises made by developers of several apps from our data set (see also Table ~\ref{table:broken_promises}). We disclosed our findings to those developers to ensure these inconsistencies can be addressed promptly (see \S~\ref{sec:disclosure} for a further discussion).

%% file: assets/tables/table_broken_promises.tex

\begin{table*}[t]
\small\centering
\begin{tabular}{lccccccc}
\hline
\textbf{App} &
  \multicolumn{1}{l}{} &
  \textbf{\begin{tabular}[c]{@{}c@{}}Privacy\\ Strategy\end{tabular}} &
  \textbf{\begin{tabular}[c]{@{}c@{}}Message\\ Content\end{tabular}} &
  \textbf{Device IDs} &
  \textbf{User IDs} &
  \textbf{Name} &
  \textbf{Phone \#} \\ \hline
\href{https://play.google.com/store/apps/details?id=com.skype.raider}{Skype} &
  (default) &
  None &
  \newmoon &
  \newmoon &
  \newmoon &
  \newmoon &
  \fullmoon \\
 &
  (secret chat) &
  E2EE &
  \fullmoon &
  \newmoon &
  \newmoon &
  \newmoon &
  \fullmoon \\ \hline
\multicolumn{2}{l}{\href{https://play.google.com/store/apps/details?id=com.snapchat.android}{Snapchat}} &
  E2EE &
  \fullmoon &
  \fullmoon &
  \newmoon &
  \newmoon &
  \fullmoon \\ \hline
\multicolumn{2}{l}{\href{https://play.google.com/store/apps/details?id=com.viber.voip}{Viber}} &
  Push-to-Sync &
  \fullmoon &
  \newmoon &
  \newmoon &
  \fullmoon &
  \newmoon \\ \hline
\multicolumn{2}{l}{\href{https://play.google.com/store/apps/details?id=jp.naver.line.android}{LINE}} &
  E2EE &
  \fullmoon &
  \fullmoon &
  \fullmoon &
  \newmoon &
  \fullmoon \\ \hline
\multicolumn{2}{l}{\href{https://play.google.com/store/apps/details?id=com.discord}{Discord}} &
  None &
  \newmoon &
  \fullmoon &
  \newmoon &
  \newmoon &
  \fullmoon \\ \hline
\multicolumn{2}{l}{\href{https://play.google.com/store/apps/details?id=com.tencent.mm}{WeChat}} &
  None &
  \newmoon &
  \fullmoon &
  \newmoon &
  \newmoon &
  \fullmoon \\ \hline
\multicolumn{2}{l}{\href{https://play.google.com/store/apps/details?id=com.juphoon.justalk}{JusTalk}} &
  None &
  \newmoon &
  \fullmoon &
  \newmoon &
  \newmoon &
  \fullmoon \\ \hline
\multicolumn{2}{l}{\href{https://play.google.com/store/apps/details?id=com.safeum.android}{SafeUM}} &
  E2EE &
  \fullmoon &
  \fullmoon &
  \newmoon &
  \fullmoon &
  \fullmoon \\ \hline
\multicolumn{2}{l}{\href{https://play.google.com/store/apps/details?id=com.yallatech.yallachat}{YallaChat}} &
  E2EE &
  \fullmoon &
  \fullmoon &
  \newmoon &
  \newmoon &
  \fullmoon \\ \hline
\multicolumn{2}{l}{\href{https://play.google.com/store/apps/details?id=com.is.core.app}{Comera}} &
  Push-to-Sync &
  \fullmoon &
  \fullmoon &
  \newmoon &
  \fullmoon &
  \newmoon \\ \hline
\multicolumn{2}{l}{\href{https://play.google.com/store/apps/details?id=com.wire}{Wire}} &
  Push-to-Sync &
  \fullmoon &
  \newmoon &
  \newmoon &
  \fullmoon &
  \fullmoon \\ \hline
\end{tabular}
\caption{This table contains all analyzed apps, for which we observed personal information leakage to FCM servers in the process of delivering push notifications. The specific observed category of data is indicated by \newmoon~(evidence) and \fullmoon~(no evidence).}
\label{table:broken_promises}
\end{table*}

%% file: paper/5_results.tex
\section{Results}
\label{sec:results}



We present findings from our analysis of secure messaging apps, including the personal information observed being shared with Google's FCM servers and the mitigation strategies employed by apps to prevent such leakage. Additionally, we analyzed statements made by app developers to determine whether they make any privacy or security guarantees and whether they disclose the sharing of personal information for push notifications.\footnote{Supplemental materials are available at \url{https://github.com/blues-lab/fcm-app-analysis-public}.} 

\subsection{App Analysis}
We found that almost all analyzed applications used FCM. Of the popular secure messaging apps that we identified, \FCMApps{} of \apps{} apps relied on FCM to deliver push notifications to users. One exception among those apps was Briar messenger, which prompted the user to enable unrestricted battery usage, allowing the app to poll for new messages in the background. (Several other apps in our dataset also prompted us to enable unrestricted battery usage, however, those apps still relied on FCM.) Since our study focuses on FCM, we excluded Briar and analyzed only those applications that relied on FCM to deliver push notifications.

Of the \FCMApps{} apps we analyzed, \leakedAny{} included personal information in data sent to Google via FCM such that that data was visible to Google. All 11 apps leaked message metadata, including device and app identifiers (\leakedDeviceIDs{} apps), user identifiers (\leakedUserIDs{} apps), the sender's or recipient's name (\leakedNames{} apps), and phone numbers (\leakedPhones{} apps). More alarmingly, we observed \leakedContents{} apps---which have cumulative installs in excess of one billion---leak message contents. We present information about the observed practices in Table~\ref{table:broken_promises}.

It is worth noting that not all of the observed behaviors here are necessarily \textit{undisclosed sharing}. Undisclosed sharing occurs when data we observed being shared from our static and/or dynamic analysis was not disclosed in the privacy disclosures we analyzed. Whether the observed behaviors do constitute undisclosed sharing depends on the findings from our privacy disclosure analysis, discussed below (\S \ref{sec:privacyDisclosures}).

\input{assets/tables/table_all_apps}

\subsection{Mitigation Strategies}
Of the \mitigationApps{} apps that did not send message contents to Google.\footnote{Skype used e2e encryption to protect message contents only in secret chats, which is not the default.} our static analysis revealed two general mitigation strategies described below: end-to-end encryption and push-to-sync. Ultimately, we observed \completeMitigationApps{} apps out of \mitigationApps{} employ either end-to-end encryption or push-to-sync strategies to prevent leaking any personal information to Google via FCM. The remaining \incompleteMitigationApps{} apps still leaked metadata, but not the message contents. See Table~\ref{table:all_apps} for more information.

\input{assets/figures/signal_payload}

\mysubsubsection{End-to-End Encryption}
\label{sec:strategy-e2ee}
We determined that \etoeApps{} apps employed an end-to-end encryption strategy to prevent privacy leakage to Google via FCM. In this strategy, when the user launches the app for the first time, the app provisions a keypair and does a secure key exchange between the user’s device and the app’s server. The app will then develop a session key that it can use to decrypt messages from the server. The server encrypts messages it sends using the session key before it goes to FCM.

As depicted in Table~\ref{table:all_apps}, of the \etoeApps{} apps that utilized the end-to-end encryption (e2e) strategy, only \etoeCompleteApps{} (Facebook Messenger, Telegram, Session, and KakaoTalk) did not leak \textit{any} personal information to Google via FCM. The remaining \etoePartialApps{} (Snapchat, SafeUM, YallaChat, and LINE) still leaked metadata, including user identifiers (\etoeLeakedUserIDs{} apps) and names (\etoeLeakedNames{} apps).

\mysubsubsection{Push-to-Sync}
\label{sec:strategy-push2sync}
We observed \ptosApps{} apps employ a push-to-sync strategy to prevent privacy leakage to Google via FCM. In this mitigation strategy, apps send an empty (or almost empty) push notification to FCM. Some apps, such as Signal, send a push notification with no data (aside from the fields that Google sets; see Figure~\ref{fig:signal}). Other apps may send an identifier (including, in some cases, a phone number). This push notification tells the app to query the app server for data, the data is retrieved securely by the app, and then a push notification is populated on the client side with the unencrypted data. In these cases, the only metadata that FCM receives is that the user received some message or messages, and when that push notification was issued. Achieving this requires sending an additional network request to the app server to fetch the data and keeping track of identifiers used to correlate the push notification received on the user device with the message on the app server.

As detailed in Table~\ref{table:all_apps}, only \ptosCompleteApps{} (Whatsapp, Signal, Threema, Element, and Kik) did not leak any personal information to Google. The remaining \ptosPartialApps{} (Viber, Wire, and Comera) leaked metadata, including user identifiers (all \ptosLeakedUserIDs{} apps), device and app identifiers (\ptosLeakedDeviceIDs{} apps), and phone numbers (\ptosLeakedPhones{} apps).

\input{assets/figures/justalk_payload}

\subsection{Privacy Disclosure Analysis}
\label{sec:privacyDisclosures}
We analyzed privacy disclosures for the \leakedAny{} apps that included personal information in the push notifications sent via Google's FCM. One of our aims was to determine whether a user could reasonably conclude that the app guarantees the security and privacy of their personal information based on the information presented by the app vendor in their Play Store description, official website, marketing and promotional materials, and other documentation. Table~\ref{table:disclosures} provides details for each app.

\mysubsubsection{Marketing Claims}
First, we discuss the \leakedContents{} apps that leaked the actual contents of push notification messages: Skype, WeChat, Discord, and JusTalk. We found that out of these four apps, only JusTalk claimed to be end-to-end secure, stating: ``All users' personal information (including calling and messaging data) is end-to-end encrypted and is split into multiple random paths which ensure it can't be monitored or saved by servers. Moreover, all the personal data is never shared with any third party. Enjoy safe and free calls''~\cite{jusTalkStatement}. Nevertheless, we clearly observed the contents of our messages being sent without end-to-end encryption via FCM's servers while delivering push notifications (see Figure~\ref{fig:justalk}).

Although the three remaining apps do not claim to employ end-to-end encryption, both WeChat and Discord made statements about their concern for privacy. For instance, WeChat said in their Play Store description: ``- BETTER PRIVACY: Giving you the highest level of control over your privacy, WeChat is certified by TRUSTe'' \cite{weChatDescription}. Although Skype does not reference secure messaging for their normal (default) chat functionality, they promise that ``Skype private conversations uses the industry standard Signal Protocol, allowing you to have end-to-end encrypted Skype audio calls, send text messages, image, audio, and video files''~\cite{skypeStatement}. Although we did not observe the content of the message being leaked when testing Skype's private conversation feature, we still observed the app leaking device IDs, user IDs, and names via Google's FCM. 

For the remaining \notLeakedContents{} apps that did not leak message contents, we observed each of these apps make claims that could lead users to believe that the apps do not share any personal information with anyone and, except for Snapchat, claimed to be end-to-end encrypted. For instance, SafeUM messenger put it plainly: ``[w]e never share your data with anyone. Never''~\cite{safeumPolicy}.

\input{assets/tables/table_disclosures}

\mysubsubsection{Privacy Policies}
We additionally read each privacy policy to understand whether developers disclosed the sharing of personal information for the purposes of providing push notifications. We found that all \leakedAny{} apps that shared personal information with Google's FCM servers stated that personal user data may be shared with service providers (such as FCM) for the purpose of app functionality. However, only two apps (JusTalk and YallaChat) enumerated the types of personal information shared with such service providers, which did not cover the types of information we observed being shared, namely user IDs and names (for both apps) and message contents (for JusTalk, as discussed above). Furthermore, three apps (Viber, WeChat and Comera) did not specify which companies serve as their service providers. Out of the remaining 8 apps, only 4 mentioned Google in the context of push notifications and/or FCM.

Given that only YallaChat included information about the types of data shared with Google's FCM, we were unable to determine whether the specific data types we observed being shared would be covered by these statements or not. For instance, Viber's privacy policy stated, without giving any specifics: ``[w]e may disclose your Personal Information to a contractor or service provider for a business purpose. The types of personal information we share for a business purpose, vary, depending on the purpose and the function provided by the third party to whom we disclose such information''~\cite{viberPolicy}. While these statements may technically address personal data sharing in the context of push notifications, they do not meaningfully inform users about \textit{what} information pertaining to them is being shared and \textit{with whom}.

%% file: assets/tables/table_all_apps.tex

\begin{table*}[t]
\small\centering
\begin{tabular}{llcccr}
\hline
\multicolumn{1}{c}{\textbf{App}}                                                           & \textbf{Version} & \textbf{Uses FCM?} & \textbf{\begin{tabular}[c]{@{}c@{}}Privacy\\ Strategy\end{tabular}}         & \textbf{\begin{tabular}[c]{@{}c@{}}Observed\\ Data Leakage\end{tabular}} & \multicolumn{1}{c}{\textbf{\begin{tabular}[c]{@{}c@{}}Min Installs\\ (millions)\end{tabular}}} \\ \hline
\href{https://play.google.com/store/apps/details?id=com.facebook.orca}{Facebook Messenger} & v403.1.0.17.106  & \newmoon           & e2ee                                                                        & \Square                                                                     & 5,000                                                                                          \\ \hline
\href{https://play.google.com/store/apps/details?id=com.whatsapp}{WhatsApp}                & v2.23.12.78      & \newmoon           & Push-to-Sync                                                                & \Square                                                                     & 5,000                                                                                          \\ \hline
\href{https://play.google.com/store/apps/details?id=com.skype.raider}{Skype}               & v8.93.0.408      & \newmoon           & \begin{tabular}[c]{@{}c@{}}\textbf{none} (default)\\ e2ee (secret chat)\end{tabular} & \XBox                                                                       & 1,000                                                                                          \\ \hline
\href{https://play.google.com/store/apps/details?id=com.snapchat.android}{Snapchat}        & v12.28.0.22      & \newmoon           & e2ee                                                                        & \XBox                                                                       & 1,000                                                                                          \\ \hline
\href{https://play.google.com/store/apps/details?id=org.telegram.messenger}{Telegram}      & v9.4.4           & \newmoon           & e2ee                                                                        & \Square                                                                     & 1,000                                                                                          \\ \hline
\href{https://play.google.com/store/apps/details?id=com.viber.voip}{Viber}                 & v19.4.0.0        & \newmoon           & Push-to-Sync                                                                & \XBox                                                                       & 1,000                                                                                          \\ \hline
\href{https://play.google.com/store/apps/details?id=jp.naver.line.android}{LINE}           & v13.4.2          & \newmoon           & e2ee                                                                        & \XBox                                                                       & 500                                                                                            \\ \hline
\href{https://play.google.com/store/apps/details?id=com.discord}{Discord}                  & v172.24          & \newmoon           & \textbf{none}                                                                        & \XBox                                                                       & 100                                                                                            \\ \hline
\href{https://play.google.com/store/apps/details?id=com.kakao.talk}{Kakao Talk}            & v10.0.7          & \newmoon           & e2ee                                                                        & \Square                                                                     & 100                                                                                            \\ \hline
\href{https://play.google.com/store/apps/details?id=kik.android}{Kik}                      & v15.50.1.27996   & \newmoon           & Push-to-Sync                                                                & \Square                                                                     & 100                                                                                            \\ \hline
\href{https://play.google.com/store/apps/details?id=org.thoughtcrime.securesms}{Signal}    & v6.11.7          & \newmoon           & Push-to-Sync                                                                & \Square                                                                     & 100                                                                                            \\ \hline
\href{https://play.google.com/store/apps/details?id=com.tencent.mm}{WeChat}                & v8.0.30          & \newmoon           & \textbf{none}                                                                        & \XBox                                                                       & 100                                                                                            \\ \hline
\href{https://play.google.com/store/apps/details?id=com.juphoon.justalk}{JusTalk}          & v8.6.10          & \newmoon           & \textbf{none}                                                                        & \XBox                                                                       & 10                                                                                             \\ \hline
\href{https://play.google.com/store/apps/details?id=com.safeum.android}{SafeUM}            & v1.1.0.1548      & \newmoon           & e2ee                                                                        & \XBox                                                                       & 5                                                                                              \\ \hline
\href{https://play.google.com/store/apps/details?id=com.yallatech.yallachat}{YallaChat}    & v1.4.2           & \newmoon           & e2ee                                                                        & \XBox                                                                       & 5                                                                                              \\ \hline
\href{https://play.google.com/store/apps/details?id=org.briarproject.briar.android}{Briar} & v1.4.23          & \fullmoon          & Polling                                                                     & \Square                                                                     & 1                                                                                              \\ \hline
\href{https://play.google.com/store/apps/details?id=com.is.core.app}{Comera}               & v4.0.1           & \newmoon           & Push-to-Sync                                                                & \XBox                                                                       & 1                                                                                              \\ \hline
\href{https://play.google.com/store/apps/details?id=im.vector.app}{Element}                & v1.5.22          & \newmoon           & Push-to-Sync                                                                & \Square                                                                     & 1                                                                                              \\ \hline
\href{https://play.google.com/store/apps/details?id=network.loki.messenger}{Session}       & v1.16.7          & \newmoon           & e2ee                                                                        & \Square                                                                     & 1                                                                                              \\ \hline
\href{https://play.google.com/store/apps/details?id=ch.threema.app}{Threema}               & v5.0.6           & \newmoon           & Push-to-Sync                                                                & \Square                                                                     & 1                                                                                              \\ \hline
\href{https://play.google.com/store/apps/details?id=com.wire}{Wire}                        & v3.82.38         & \newmoon           & Push-to-Sync                                                                & \XBox                                                                       & 1                                                                                              \\ \hline
\multicolumn{1}{r}{}                                                                       &                  &                    & \textbf{}                                                                   & TOTAL installs                                                              & \textbf{15,026}                                                                                \\ \cline{5-6} 
\end{tabular}
\caption{Our data set of analyzed apps. Usage of Firebase Cloud Messaging (FCM) is indicated by \newmoon (does use) and \fullmoon (does not use). Whether or not an app leaked personal information to FCM is indicated by \Square (no evidence) and \XBox (evidence). See Table~\ref{table:broken_promises} for details on which personal data is leaked by apps marked with \XBox. Apps are sorted by minimum install count and alphabetically by app name.}
\label{table:all_apps}
\end{table*}

%% file: assets/figures/signal_payload.tex
\begin{figure}[t]
  \centering
  \fbox{\includegraphics[width=0.75\columnwidth]{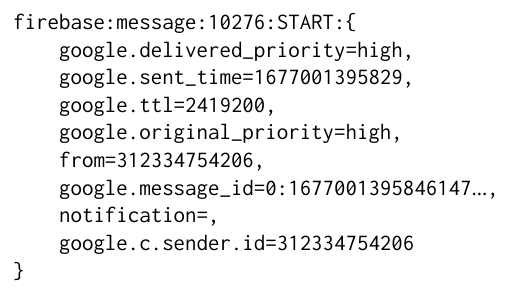}}
  \caption{Example payload from within the \texttt{RemoteMessage} object received by the Signal app. Note the empty notification field, indicating the correct usage of the \textit{push-to-sync} notification strategy.}
  \label{fig:signal}
\end{figure}

%% file: assets/figures/justalk_payload.tex
\begin{figure}[h!]
  \centering
  \fbox{\includegraphics[width=0.9\columnwidth]{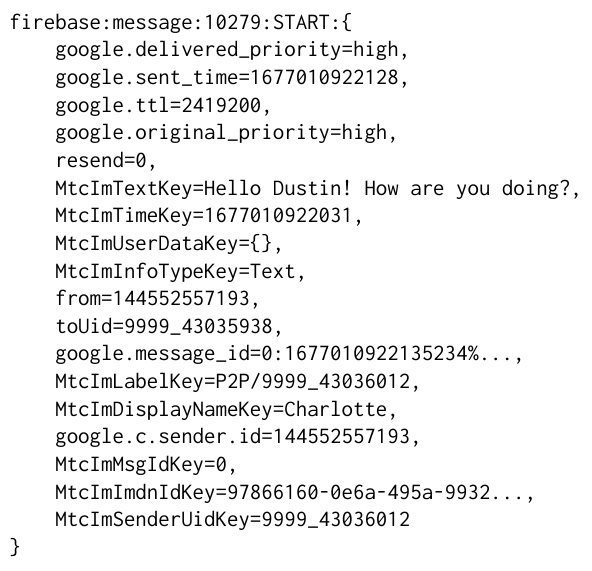}}
  \caption{Payload contained inside the \texttt{RemoteMessage} object received by JusTalk. Note the \texttt{MtcImTextKey} and \texttt{MtcImDisplayNameKey}, which contain the unencrypted message contents and username, respectively.}
  \label{fig:justalk}
\end{figure}

%% file: assets/tables/table_disclosures.tex
\begin{table*}[t]
\small\centering
\begin{tabular}{lcccccc}
\hline
\textbf{App} &
  \multicolumn{1}{l}{} &
  \textbf{E2EE} &
  \textbf{\begin{tabular}[c]{@{}c@{}}S/P \end{tabular}} &
  \textbf{\begin{tabular}[c]{@{}c@{}}Discloses \\ PI Sharing\end{tabular}} &
  \textbf{\begin{tabular}[c]{@{}c@{}}Discloses\\ Companies\end{tabular}} &
  \textbf{\begin{tabular}[c]{@{}c@{}}Discloses\\ Shared PI\end{tabular}} \\ \hline
\href{https://play.google.com/store/apps/details?id=com.skype.raider}{Skype}         & (default)            & \fullmoon & \fullmoon & \newmoon & \newmoon   &  \fullmoon \\
                                                                                     & (secret chat)        & \newmoon   & \newmoon   & \newmoon & \newmoon   & \fullmoon \\ \hline
\multicolumn{2}{l}{\href{https://play.google.com/store/apps/details?id=com.snapchat.android}{Snapchat}}     & \fullmoon & \newmoon   & \newmoon & \newmoon   & \fullmoon \\ \hline
\multicolumn{2}{l}{\href{https://play.google.com/store/apps/details?id=com.viber.voip}{Viber}}              & \newmoon   & \newmoon   & \newmoon & \fullmoon & \fullmoon \\ \hline
\multicolumn{2}{l}{\href{https://play.google.com/store/apps/details?id=jp.naver.line.android}{LINE}}        & \newmoon   & \newmoon   & \newmoon & \newmoon   & \fullmoon \\ \hline
\multicolumn{2}{l}{\href{https://play.google.com/store/apps/details?id=com.discord}{Discord}}               & \fullmoon & \newmoon   & \newmoon & \newmoon   & \fullmoon \\ \hline
\multicolumn{2}{l}{\href{https://play.google.com/store/apps/details?id=com.tencent.mm}{WeChat}}             & \fullmoon & \newmoon   & \newmoon & \fullmoon & \fullmoon \\ \hline
\multicolumn{2}{l}{\href{https://play.google.com/store/apps/details?id=com.juphoon.justalk}{JusTalk}}       & \newmoon   & \newmoon   & \newmoon & \newmoon   & \newmoon   \\ \hline
\multicolumn{2}{l}{\href{https://play.google.com/store/apps/details?id=com.safeum.android}{SafeUM}}         & \newmoon   & \newmoon   & \newmoon & \newmoon   & \fullmoon \\ \hline
\multicolumn{2}{l}{\href{https://play.google.com/store/apps/details?id=com.yallatech.yallachat}{YallaChat}} & \newmoon   & \newmoon   & \newmoon & \newmoon   & \newmoon   \\ \hline
\multicolumn{2}{l}{\href{https://play.google.com/store/apps/details?id=com.is.core.app}{Comera}}            & \newmoon   & \newmoon   & \newmoon & \fullmoon & \fullmoon \\ \hline
\multicolumn{2}{l}{\href{https://play.google.com/store/apps/details?id=com.wire}{Wire}}                     & \newmoon   & \newmoon   & \newmoon & \newmoon   & \fullmoon \\ \hline
\end{tabular}
\caption{This table contains information about the disclosures made by developers of apps, for which we observed information leakage to FCM. \newmoon~indicates that we found evidence (or \fullmoon~if not) for each of the following statements: \textbf{(E2EE)} developer states the app uses end-to-end encryption, (\textbf{S/P}) developer makes security or privacy-specific claims in the Google Play Store description or on their official websites, (\textbf{discloses PI sharing}) developer discloses in their privacy policy the sharing of personal information to service providers for app functionality purposes, (\textbf{discloses companies}) if the disclosure includes names of companies and (\textbf{discloses shared PI}) if the disclosure includes specific types of personal information.}
\label{table:disclosures}
\end{table*}

%% file: paper/6_discussion.tex
\section{Discussion}
\label{sec:discussion}
The democratization of mass communications via the Internet has created a new paradigm in which anyone can have a platform to send a message. Consequently, anyone can now become a software engineer and distribute software worldwide. By and large, this is a good thing. However, it raises issues of professional responsibility that have long been addressed by other more mature branches of engineering. In most jurisdictions, one cannot simply decide to become a civil engineer and erect a multi-story building. Due to the inherent safety risks---to the individual, neighbors, and society---most jurisdictions require that plans be submitted for approval. In granting that approval, the plans are first checked for conformance with building codes, which have been set (and periodically revised) by professional societies with deep expertise. Once plans are approved, multiple levels of oversight still occur: at various steps during construction, building inspectors confirm that both the plans have been followed and that no other safety issues have been identified. Moreover, after construction has been completed, governments are empowered to continually monitor for code violations, going so far as to condemn structures that pose safety hazards. Of course, there is a reason for this oversight: building codes are written in blood.

In the past decade or two, software engineering as a discipline has only just begun to reckon with the complex sociotechnical issues relating to harm and liability. While the collapse of a building is likely to be more lethal than the inappropriate exfiltration of sensitive user information, the latter may still pose risks to user safety---even lethal ones. We chose to examine secure messaging apps in this study because they can often embody these risks: online messaging apps are increasingly being used by activists living in oppressive regimes~\cite{tufekci2017twitter}, who may find themselves in serious jeopardy if their communications are inappropriately revealed. In this specific instance, the inappropriate disclosure of users' communication and metadata does not require malice on the part of a service provider for harm to come to the user. By nature of such data collection, the service provider exposes the user to legal processes: this may result in data the user legitimately did not believe to exist coming into the hands of governments and private actors. We emphasize that this risk is not merely theoretical; as previously noted, U.S. Senator Ron Wyden published a letter that confirms that government agencies do, in fact, collect user information by demanding push notification records from Google and Apple~\cite{wydenLetter}.

Our analysis found that several prevalent secure messaging apps---which imply that they will not share certain information with third parties---do indeed share that information in plaintext with Google via FCM (see Table~\ref{table:broken_promises}). We found evidence of undisclosed data leakage to FCM in apps that account for over 2 billion installs. Users of these apps are likely unaware of these data leakages: some of the privacy disclosures made by these apps often explicitly promise \emph{not} to share such personal information with third parties, whereas others were so vaguely written that it was unclear whether these behaviors are being disclosed (and how they might comport in consumers' minds with the companies' marketing materials that imply messaging data will be kept private). Consequently, consumers may have a false sense of security when using these apps for communicating. The undisclosed leakage of communication contents can harm users and potentially even innocent bystanders who may be mentioned in communications.

\subsection{Recommendations}
Just as a contractor or owner-builder is ultimately responsible for the adherence to local building codes and the risks associated with deviations from them, software developers publishing apps for public usage are responsible for the behaviors of those apps. This responsibility includes verifying that third-party components function as expected and that the ultimate behavior of the app is in accordance with platform guidelines, the developer's disclosures, and applicable laws/regulations. The use of these third-party components is not unique to software engineering: other branches of engineering generally involve complex supply chains, yet there is often a great deal of oversight. When Airbus builds a plane, they may use engines from Rolls-Royce or electronics from Siemens; but in addition to simply specifying the specifications and tolerances that Airbus expects these components to conform to, they nonetheless validate those third-party components by launching chickens at them at 600+ km/h (amongst other validation tests)~\cite{ChickenGun}. Such integration validations rarely exist for software \textit{in practice}, despite being recommended for nearly half a century now~\cite{Brooks1975}. Indeed, while we have no reason to believe that misleading or confusing security and privacy claims are the result of malice, we believe that the poor privacy practices that we document in this paper could have been discovered and mitigated by the developers had they inspected the traffic sent and received by their applications during quality assurance processes. Thus, we offer recommendations to different stakeholders on ways to address the identified security and privacy issues.

\subsubsection{App Developers} As the parties ultimately responsible for their apps, app developers should perform the type of dynamic analysis that we performed in this study as part of each and every release cycle. This will help to ensure that users' personal data flows in accordance with reasonable expectations, applicable laws and regulations, as well as platform policies. However, the best way to ensure that push notifications do not leak sensitive user information is to avoid sending sensitive user information via FCM in the first place. We argue that developers should implement the push-to-sync approach: the developer’s server should only send the app a unique notification ID via FCM, which can then be used to fetch the notification content from the developer’s servers securely. Several developers correctly used the push-to-sync approach, which resulted in no personal data being leaked by those apps. Others should adopt this architecture in their apps.



\subsubsection{Platforms and SDK Providers} At the same time, platform owners and SDK providers are well-positioned to identify and correct issues in their tools and highlight security and privacy risks in their documentation. For its part, Google provides an API that results in developers systematically making very similar privacy mistakes. This is not helped by Google's guidance, which instructs developers to ``send as much data as possible in the FCM payload,'' and that if they want to do so securely, they must use an additional library~\cite{androidDevBlog}. This guidance departs from Google's own data minimization and secure-by-default principles~\cite{GoogleSecurity} and recommendations from other push notification providers, such as Apple~\cite{applePush}. 


We argue that the availability of usable, secure push notifications libraries, including Google’s Capillary~\cite{capillary}, does not solve the underlying problem. Developers generally trust Google's security practices and are largely unaware of the risk of personal information leakage via push notifications. Furthermore, under current regulatory regimes, Google is not obligated to do anything about this: they provide a free API for developers, and despite the fact that using it to send messages \textit{securely} admittedly takes additional non-obvious steps, there are no legal requirements that Google---or any other SDK provider---provide a secure-by-default API\@. Furthermore, as mentioned previously, Android app developers are effectively required to use Google's FCM to send push notifications for battery consumption reasons. We argue, therefore, that real-world change will require either applying regulatory pressure or other market-corrective forces on platform owners to enforce privacy-by-design principles for critical SDKs in the software supply chain, such as Google's FCM. Such a change would improve the privacy and security of nearly all Android apps, because the use of FCM to deliver push notifications on Android is nearly universal.


The use of these types of APIs also represents the classic usable security problem (wherein application developers are the ``user''): the user is not qualified to be making the decisions that are forced upon them, whereas those forcing them to make these decisions are in a much better position to make those decisions on the users' behalf. Prior research shows that developers, despite being the party ultimately responsible for the behaviors of their software, are woefully unprepared to make these types of decisions~\cite{alomar2022developers,acar2017comparing}. And thus, we are faced with a situation in which the parties most equipped to fix these types of problems (e.g., by creating more usable documentation that highlights security and privacy risks, making SDK settings secure by default, proactively auditing how their services are used in practice, etc.) are not incentivized to do so, whereas the parties who are ultimately responsible are generally incapable and do not understand their risks or responsibilities. As a result, this is fundamentally an economics problem concerning misaligned incentives~\cite{anderson2001}: in a perfect world, the responsibility for handling users' data responsibly would be placed upon those according to their abilities, shifted from those according to their needs~\cite{Marx1875}. This is not the world in which we currently live.

Yet, things are improving. In recent years, the U.S. Government has promoted the strategy of shifting the burden of software security away from individuals, small businesses, and local governments and onto the organizations that are most capable and best-positioned to reduce risks~\cite{whiteHouseStrategy}. In line with this initiative, the U.S. Cybersecurity and Infrastructure Security Agency (CISA) and 17 U.S. and international partners published an update in August 2023 to joint guidance for implementing secure-by-design principles~\cite{cisaReport}. One secure product development practice, in particular, highlights the need to provide secure defaults for developers by ``providing safe building blocks...known as `paved roads' or `well-lit paths.'{}'' We believe that push notification providers can similarly apply privacy-by-design principles~\cite{pallas2024privacy} to safeguard the privacy of users who cannot easily manage the risks.

Without correctly aligned incentives to motivate platforms and SDK providers to make their systems secure by default (including documentation that highlights security and privacy risks), developers will continue to be placed in this position and will continue to consistently make these types of mistakes. Thus, until software engineering becomes a more mature field with formalized oversight, validation, disclosure, and auditing procedures, these types of errors will proliferate, leaving end users at risk.

%% file: paper/6a_disclosure.tex
\section{Responsible Disclosure}
\label{sec:disclosure}

Responsible disclosure is a critical component of security and privacy research. We reported our substantive findings to the \leakedAny{} app developers who leaked at least one personal data type to Google's FCM service. We tried contacting the developers via various contact methods, including formal bug bounty programs, emailing security teams, or failing that, general support contacts. The app developers for whom we could find contact information were sent a summary report on or before June 7, 2024. We received an acknowledgment of our email from 5 developers of the \leakedAny{} we contacted.

At the time of publication, the remaining 6 app developers to whom we disclosed our findings had not replied; discussions are ongoing with several companies regarding how they should fix the identified issues. We look forward to continue engaging in productive conversations to help developers understand how to adapt their push message architectures to better protect user privacy.

%% file: paper/7_limitations.tex
\section{Limitations}
\label{sec:limitations}

Many apps beyond secure messaging apps might send private data through push notifications. Our study only focused on secure messaging apps because most of them claim to focus on user privacy, thus, they would be among the most likely apps to take proactive steps to prevent the leakage of user data to FCM (and presumably users of these apps are more likely to believe that their communications are secure). We suspect that privacy leakage via Google FCM may be even more prevalent within apps in other contexts. Future work should look at both less popular secure messaging apps and apps in other contexts to observe to what extent, if at any, they mitigate the leakage of sensitive personal data to Google via FCM.

We also performed our analysis using an older Pixel 3a device running Android 12. We are unaware of any substantial changes in Android 13 and 14 that would have a material impact on our observed findings. Our device supported security updates and the installation of all the apps that we analyzed for this research. We ran these apps and received push notifications from FCM without observing any undesirable impact on app performance. Furthermore, at the time we began our analysis in early 2023, the majority of users (more than 85\%) used Android version 12 or below~\cite{androidMarketShare}. While most people who use a mobile phone use an Android device, iOS also has a significant share of the mobile phone market and tends to bill itself as having more privacy-preserving practices. Future work can also explore whether private user data is leaked to Apple or other third parties via the push notification infrastructure available to developers in the iOS ecosystem. 

We looked specifically at privacy leakage through push notifications that rely on FCM. As far as we know, FCM is also used in other applications, on Android and beyond; how this fact affects privacy leakage across other applications is not well understood. Future work could investigate the privacy implications of FCM across those applications. Within the Android ecosystem, there may exist other patterns or tools provided by Google or by other popular third-party libraries that also incur unexpected privacy leakage. Future work could look for such patterns beyond the Android platform, such as iOS, and identify how other ecosystem players like Apple and Google can craft a more trustworthy ecosystem to provide more privacy-preserving defaults to the broadest base of users.

\begin{quote}
{\it ``The personal and social consequences of any medium---that is, of any extension of ourselves---result from the new scale that is introduced into our affairs by each extension of ourselves, or by any new technology''}

\hfill ---Marshall McLuhan~\cite{McLuhan1964}.
\end{quote}

%% file: paper/appendix.tex
\newpage\appendix

\section{Data Types}
\label{sec:datatypes}
Table~\ref{table:data_types} enumerates the data types that we searched for during our analysis of Android apps. Google defines and uses these data types to populate the information presented to users in the form of privacy labels in the app's listing on Google Play Store~\cite{googlePrivacyLabels}.

\input{assets/tables/table_data_types}

\section{Code Analysis Workflow}
\label{sec:workflow}
We used this set of questions to analyze the source code of apps in our data set. These questions can also assist with data flow mapping, or in other words, tracing data contained in a push notification from its creation until the notification is displayed to the user.

\begin{itemize}
    \item Does the app's \texttt{AndroidManifest.xml} register a service that \texttt{extends FirebaseMessagingService}?
    \item Locate the Java .java (or Kotlin .kt) source file corresponding to the registered service.
    \item Which FCM methods (e.g., \texttt{onMessageReceived()},\\ \texttt{onNewToken()}, etc.) does the service override? 
    \item The \texttt{onMessageReceived()} method gets invoked when the client app receives an FCM push notification. Does the service override \texttt{onMessageReceived()} method?
    \item Data payload contained in an FCM push notification can be accessed by calling \texttt{remoteMessage.getData()}. Does the \texttt{onMessageReceived()} method invoke \texttt{getData()} on its argument of type \texttt{RemoteMessage}?
    \item Is there any indication that \texttt{RemoteMessage} contains sensitive data, based on the names of the keys or logging?
    \item Trace the code execution from the \texttt{onMessageReceived()} method until the message is displayed to the user.
    \item Does \texttt{RemoteMessage} get passed as a parameter to any function?
    \item What mechanisms (if any) are in place to ensure that notification contents do not get leaked to Google's FCM server?
\end{itemize}

%% file: assets/tables/table_data_types.tex
\begin{table}[t]
\centering
\resizebox{0.8\columnwidth}{!}{%
\begin{tabular}{ll}
\hline
\multicolumn{1}{c}{\textbf{Data Type}} &
  \textbf{Description} \\ \hline
\textbf{\begin{tabular}[c]{@{}l@{}}Device or \\ other IDs\end{tabular}} &
  \begin{tabular}[c]{@{}l@{}}Identifiers that relate to an individual \\ device, browser or app. For example, \\ an IMEI number, MAC address, Wi-\\ devine Device ID, Firebase installa-\\ tion ID, or advertising identifier.\end{tabular} \\ \hline
\textbf{User IDs} &
  \begin{tabular}[c]{@{}l@{}}Identifiers that relate to an identifiable \\ person. For example, an account ID, \\ account number, or account name.\end{tabular} \\ \hline
\textbf{Name} &
  \begin{tabular}[c]{@{}l@{}}How a user refers to themselves, such \\ as their first or last name, or nickname.\end{tabular} \\ \hline
\textbf{\begin{tabular}[c]{@{}l@{}}Phone \\ number\end{tabular}} &
  A user’s phone number. \\ \hline
\textbf{Messages} &
  \begin{tabular}[c]{@{}l@{}}Any other types of messages. For \\ example, instant messages or chat \\ content.\end{tabular} \\ \hline
\end{tabular}%
}
\caption{Google Play Store's data types applicable to our study. Note that Google refers to the `Messages' data type as `Other in-app messages.'}
\label{table:data_types}
\end{table}

%% file: main.bbl

\begin{thebibliography}{105}


\ifx \showCODEN    \undefined \def \showCODEN     #1{\unskip}     \fi
\ifx \showDOI      \undefined \def \showDOI       #1{#1}\fi
\ifx \showISBNx    \undefined \def \showISBNx     #1{\unskip}     \fi
\ifx \showISBNxiii \undefined \def \showISBNxiii  #1{\unskip}     \fi
\ifx \showISSN     \undefined \def \showISSN      #1{\unskip}     \fi
\ifx \showLCCN     \undefined \def \showLCCN      #1{\unskip}     \fi
\ifx \shownote     \undefined \def \shownote      #1{#1}          \fi
\ifx \showarticletitle \undefined \def \showarticletitle #1{#1}   \fi
\ifx \showURL      \undefined \def \showURL       {\relax}        \fi
\providecommand\bibfield[2]{#2}
\providecommand\bibinfo[2]{#2}
\providecommand\natexlab[1]{#1}
\providecommand\showeprint[2][]{arXiv:#2}

\bibitem[Acar et~al\mbox{.}(2017)]%
        {acar2017comparing}
\bibfield{author}{\bibinfo{person}{Yasemin Acar}, \bibinfo{person}{Michael Backes}, \bibinfo{person}{Sascha Fahl}, \bibinfo{person}{Simson Garfinkel}, \bibinfo{person}{Doowon Kim}, \bibinfo{person}{Michelle~L Mazurek}, {and} \bibinfo{person}{Christian Stransky}.} \bibinfo{year}{2017}\natexlab{}.
\newblock \showarticletitle{Comparing the usability of cryptographic apis}. In \bibinfo{booktitle}{\emph{2017 IEEE Symposium on Security and Privacy (SP)}}. IEEE, \bibinfo{pages}{154--171}.
\newblock


\bibitem[Ahmadi et~al\mbox{.}(2016)]%
        {ahmadi2016detecting}
\bibfield{author}{\bibinfo{person}{Mansour Ahmadi}, \bibinfo{person}{Battista Biggio}, \bibinfo{person}{Steven Arzt}, \bibinfo{person}{Davide Ariu}, {and} \bibinfo{person}{Giorgio Giacinto}.} \bibinfo{year}{2016}\natexlab{}.
\newblock \showarticletitle{Detecting misuse of google cloud messaging in android badware}. In \bibinfo{booktitle}{\emph{Proceedings of the 6th Workshop on Security and Privacy in Smartphones and Mobile Devices}}. \bibinfo{pages}{103--112}.
\newblock


\bibitem[AirShip(2023)]%
        {airshipStatement}
\bibfield{author}{\bibinfo{person}{AirShip}.} \bibinfo{year}{2023}\natexlab{}.
\newblock \bibinfo{title}{{Android SDK Setup}}.
\newblock \bibinfo{howpublished}{\url{https://docs.airship.com/platform/mobile/setup/sdk/android/}}.
\newblock
\newblock
\shownote{(Accessed on 10/10/2023)}.


\bibitem[Alomar and Egelman(2022)]%
        {alomar2022developers}
\bibfield{author}{\bibinfo{person}{Noura Alomar} {and} \bibinfo{person}{Serge Egelman}.} \bibinfo{year}{2022}\natexlab{}.
\newblock \showarticletitle{Developers say the darnedest things: Privacy compliance processes followed by developers of child-directed apps}.
\newblock \bibinfo{journal}{\emph{Proceedings on Privacy Enhancing Technologies}} \bibinfo{volume}{4}, \bibinfo{number}{2022} (\bibinfo{year}{2022}), \bibinfo{pages}{24}.
\newblock


\bibitem[Anderson(2001)]%
        {anderson2001}
\bibfield{author}{\bibinfo{person}{R. Anderson}.} \bibinfo{year}{2001}\natexlab{}.
\newblock \showarticletitle{Why information security is hard - an economic perspective}. In \bibinfo{booktitle}{\emph{Seventeenth Annual Computer Security Applications Conference}}. \bibinfo{pages}{358--365}.
\newblock
\urldef\tempurl%
\url{https://doi.org/10.1109/ACSAC.2001.991552}
\showDOI{\tempurl}


\bibitem[Andow et~al\mbox{.}(2019)]%
        {andow2019policylint}
\bibfield{author}{\bibinfo{person}{Benjamin Andow}, \bibinfo{person}{Samin~Yaseer Mahmud}, \bibinfo{person}{Wenyu Wang}, \bibinfo{person}{Justin Whitaker}, \bibinfo{person}{William Enck}, \bibinfo{person}{Bradley Reaves}, \bibinfo{person}{Kapil Singh}, {and} \bibinfo{person}{Tao Xie}.} \bibinfo{year}{2019}\natexlab{}.
\newblock \showarticletitle{{PolicyLint}: Investigating Internal Privacy Policy Contradictions on Google Play}. In \bibinfo{booktitle}{\emph{28th USENIX security symposium (USENIX security 19)}}. \bibinfo{publisher}{USENIX}, \bibinfo{address}{Berkeley, CA, USA}, \bibinfo{pages}{585--602}.
\newblock


\bibitem[Andow et~al\mbox{.}(2020)]%
        {andow2020actions}
\bibfield{author}{\bibinfo{person}{Benjamin Andow}, \bibinfo{person}{Samin~Yaseer Mahmud}, \bibinfo{person}{Justin Whitaker}, \bibinfo{person}{William Enck}, \bibinfo{person}{Bradley Reaves}, \bibinfo{person}{Kapil Singh}, {and} \bibinfo{person}{Serge Egelman}.} \bibinfo{year}{2020}\natexlab{}.
\newblock \showarticletitle{Actions Speak Louder than Words:{Entity-Sensitive} Privacy Policy and Data Flow Analysis with {PoliCheck}}. In \bibinfo{booktitle}{\emph{29th USENIX Security Symposium (USENIX Security 20)}}. \bibinfo{publisher}{USENIX}, \bibinfo{address}{Berkeley, CA, USA}, \bibinfo{pages}{985--1002}.
\newblock


\bibitem[Apple(2023)]%
        {AppleNotifications}
\bibfield{author}{\bibinfo{person}{Apple}.} \bibinfo{year}{2023}\natexlab{}.
\newblock \bibinfo{title}{Notifications Overview}.
\newblock \bibinfo{howpublished}{Apple Developer}.
\newblock
\newblock
\shownote{\url{https://developer.apple.com/notifications/}}.


\bibitem[{Apple}(2023)]%
        {appleTransparencyReport}
\bibfield{author}{\bibinfo{person}{{Apple}}.} \bibinfo{year}{2023}\natexlab{}.
\newblock \bibinfo{title}{{Push Token Requests}}.
\newblock \bibinfo{howpublished}{\url{https://www.apple.com/legal/transparency/push-token.html}}.
\newblock
\newblock
\shownote{(Accessed on 06/01/2024)}.


\bibitem[{Apple Inc.}(2023)]%
        {applePush}
\bibfield{author}{\bibinfo{person}{{Apple Inc.}}} \bibinfo{year}{2023}\natexlab{}.
\newblock \bibinfo{title}{{Generating a remote notification }}.
\newblock \bibinfo{howpublished}{\url{https://developer.apple.com/documentation/usernotifications/setting_up_a_remote_notification_server/generating_a_remote_notification}}.
\newblock
\newblock
\shownote{(Accessed on 10/10/2023)}.


\bibitem[Archive(2023)]%
        {waybackMachine}
\bibfield{author}{\bibinfo{person}{Internet Archive}.} \bibinfo{year}{2023}\natexlab{}.
\newblock \bibinfo{title}{{Wayback Machine}}.
\newblock \bibinfo{howpublished}{\url{https://archive.org/}}.
\newblock
\newblock
\shownote{(Accessed on 10/10/2023)}.


\bibitem[Basques and Gaunt(2023)]%
        {pushOverview}
\bibfield{author}{\bibinfo{person}{Kayce Basques} {and} \bibinfo{person}{Matt Gaunt}.} \bibinfo{year}{2023}\natexlab{}.
\newblock \bibinfo{title}{{Push notifications overview}}.
\newblock \bibinfo{howpublished}{\url{https://web.dev/articles/push-notifications-overview}}.
\newblock
\newblock
\shownote{(Accessed on 10/10/2023)}.


\bibitem[Blog(2018)]%
        {capillary}
\bibfield{author}{\bibinfo{person}{Android~Developers Blog}.} \bibinfo{year}{2018}\natexlab{}.
\newblock \bibinfo{title}{{Project Capillary: End-to-end encryption for push messaging, simplified}}.
\newblock \bibinfo{howpublished}{\url{https://android-developers.googleblog.com/2018/06/project-capillary-end-to-end-encryption.html}}.
\newblock
\newblock
\shownote{(Accessed on 10/10/2023)}.


\bibitem[Bui et~al\mbox{.}(2021)]%
        {bui2021automated}
\bibfield{author}{\bibinfo{person}{Duc Bui}, \bibinfo{person}{Kang~G Shin}, \bibinfo{person}{Jong-Min Choi}, {and} \bibinfo{person}{Junbum Shin}.} \bibinfo{year}{2021}\natexlab{}.
\newblock \showarticletitle{Automated Extraction and Presentation of Data Practices in Privacy Policies.}
\newblock \bibinfo{journal}{\emph{Proceedings on Privacy Enhancing Technologies (PoPETs)}} \bibinfo{volume}{2021}, \bibinfo{number}{2} (\bibinfo{year}{2021}), \bibinfo{pages}{88--110}.
\newblock


\bibitem[Cavallaro et~al\mbox{.}(2008)]%
        {Cavallaro2008}
\bibfield{author}{\bibinfo{person}{L. Cavallaro}, \bibinfo{person}{P. Saxena}, {and} \bibinfo{person}{R. Sekar}.} \bibinfo{year}{2008}\natexlab{}.
\newblock \showarticletitle{{On the Limits of Information Flow Techniques for Malware Analysis and Containment}}. In \bibinfo{booktitle}{\emph{Proc. of DIMVA}}. \bibinfo{publisher}{Springer-Verlag}, \bibinfo{pages}{143--163}.
\newblock
\urldef\tempurl%
\url{http://dx.doi.org/10.1007/978-3-540-70542-0_8}
\showURL{%
\tempurl}


\bibitem[Cavoukian(2009)]%
        {cavoukian2009privacy}
\bibfield{author}{\bibinfo{person}{Ann Cavoukian}.} \bibinfo{year}{2009}\natexlab{}.
\newblock \showarticletitle{Privacy by design}.
\newblock  (\bibinfo{year}{2009}).
\newblock


\bibitem[Chen et~al\mbox{.}(2015)]%
        {chen2015perplexed}
\bibfield{author}{\bibinfo{person}{Yangyi Chen}, \bibinfo{person}{Tongxin Li}, \bibinfo{person}{XiaoFeng Wang}, \bibinfo{person}{Kai Chen}, {and} \bibinfo{person}{Xinhui Han}.} \bibinfo{year}{2015}\natexlab{}.
\newblock \showarticletitle{Perplexed messengers from the cloud: Automated security analysis of push-messaging integrations}. In \bibinfo{booktitle}{\emph{Proceedings of the 22nd ACM SIGSAC Conference on Computer and Communications Security}}. \bibinfo{pages}{1260--1272}.
\newblock


\bibitem[Commission(2021)]%
        {FTCFlo2021}
\bibfield{author}{\bibinfo{person}{U.S. Federal~Trade Commission}.} \bibinfo{year}{2021}\natexlab{}.
\newblock \bibinfo{title}{Flo Health, Inc.}
\newblock
\newblock
\newblock
\shownote{\url{https://www.ftc.gov/legal-library/browse/cases-proceedings/192-3133-flo-health-inc}}.


\bibitem[Commission(2024)]%
        {FTCAvast2024}
\bibfield{author}{\bibinfo{person}{U.S. Federal~Trade Commission}.} \bibinfo{year}{2024}\natexlab{}.
\newblock \bibinfo{title}{Avast, Ltd.}
\newblock
\newblock
\newblock
\shownote{\url{https://www.ftc.gov/system/files/ftc_gov/pdf/Complaint-Avast.pdf}}.


\bibitem[{Cox, Joseph}(2023)]%
        {pushWarrant3}
\bibfield{author}{\bibinfo{person}{{Cox, Joseph}}.} \bibinfo{year}{2023}\natexlab{}.
\newblock \bibinfo{title}{{Here's a Warrant Showing the U.S. Government is Monitoring Push Notifications}}.
\newblock \bibinfo{howpublished}{\url{https://www.404media.co/us-government-warrant-monitoring-push-notifications-apple-google-yahoo/}}.
\newblock
\newblock
\shownote{(Accessed on 06/01/2024)}.


\bibitem[{Cybersecurity and Infrastructure Security Agency (CISA)}(2023)]%
        {cisaReport}
\bibfield{author}{\bibinfo{person}{{Cybersecurity and Infrastructure Security Agency (CISA)}}.} \bibinfo{year}{2023}\natexlab{}.
\newblock \bibinfo{title}{{Shifting the Balance of Cybersecurity Risk: Principles and Approaches for Secure by Design Software}}.
\newblock \bibinfo{howpublished}{\url{https://www.cisa.gov/sites/default/files/2023-10/SecureByDesign_1025_508c.pdf}}.
\newblock
\newblock
\shownote{(Accessed on 06/01/2024)}.


\bibitem[Electronics(2023)]%
        {samsungPush}
\bibfield{author}{\bibinfo{person}{Samsung Electronics}.} \bibinfo{year}{2023}\natexlab{}.
\newblock \bibinfo{title}{Samsung Push Service}.
\newblock \bibinfo{howpublished}{\url{https://play.google.com/store/apps/details?id=com.sec.spp.push}}.
\newblock
\newblock
\shownote{(Accessed on 06/01/2024)}.


\bibitem[Enck et~al\mbox{.}(2010)]%
        {Enck2010}
\bibfield{author}{\bibinfo{person}{W. Enck}, \bibinfo{person}{P. Gilbert}, \bibinfo{person}{B. Chun}, \bibinfo{person}{L.~P. Cox}, \bibinfo{person}{J. Jung}, \bibinfo{person}{P. McDaniel}, {and} \bibinfo{person}{A.~N. Sheth}.} \bibinfo{year}{2010}\natexlab{}.
\newblock \showarticletitle{{TaintDroid: An Information-flow Tracking System for Realtime Privacy Monitoring on Smartphones}}. In \bibinfo{booktitle}{\emph{Proc. of the 9th USENIX conference on Operating systems design and implementation (OSDI)}}. \bibinfo{pages}{393--407}.
\newblock


\bibitem[Fan et~al\mbox{.}(2020)]%
        {fan2020empirical}
\bibfield{author}{\bibinfo{person}{Ming Fan}, \bibinfo{person}{Le Yu}, \bibinfo{person}{Sen Chen}, \bibinfo{person}{Hao Zhou}, \bibinfo{person}{Xiapu Luo}, \bibinfo{person}{Shuyue Li}, \bibinfo{person}{Yang Liu}, \bibinfo{person}{Jun Liu}, {and} \bibinfo{person}{Ting Liu}.} \bibinfo{year}{2020}\natexlab{}.
\newblock \showarticletitle{An empirical evaluation of GDPR compliance violations in Android mHealth apps}. In \bibinfo{booktitle}{\emph{2020 IEEE 31st international symposium on software reliability engineering (ISSRE)}}. \bibinfo{publisher}{IEEE}, \bibinfo{address}{New York, NY, USA}, \bibinfo{pages}{253--264}.
\newblock


\bibitem[{Federal Trade Commision (FTC)}(2020)]%
        {ftcStatement}
\bibfield{author}{\bibinfo{person}{{Federal Trade Commision (FTC)}}.} \bibinfo{year}{2020}\natexlab{}.
\newblock \bibinfo{title}{{FTC Requires Zoom to Enhance its Security Practices as Part of Settlement}}.
\newblock \bibinfo{howpublished}{\url{https://www.ftc.gov/news-events/news/press-releases/2020/11/ftc-requires-zoom-enhance-its-security-practices-part-settlement}}.
\newblock
\newblock
\shownote{(Accessed on 01/01/2024)}.


\bibitem[Felt et~al\mbox{.}(2012)]%
        {Felt2012}
\bibfield{author}{\bibinfo{person}{A.~P. Felt}, \bibinfo{person}{E. Ha}, \bibinfo{person}{S. Egelman}, \bibinfo{person}{A. Haney}, \bibinfo{person}{E. Chin}, {and} \bibinfo{person}{D. Wagner}.} \bibinfo{year}{2012}\natexlab{}.
\newblock \showarticletitle{Android permissions: user attention, comprehension, and behavior}. In \bibinfo{booktitle}{\emph{Proceedings of the 8th Symposium on Usable Privacy and Security}} (Washington, D.C.) \emph{(\bibinfo{series}{SOUPS '12})}. \bibinfo{publisher}{ACM}, \bibinfo{address}{New York, NY, USA}, Article \bibinfo{articleno}{3}, \bibinfo{numpages}{14}~pages.
\newblock
\showISBNx{978-1-4503-1532-6}
\urldef\tempurl%
\url{https://doi.org/10.1145/2335356.2335360}
\showDOI{\tempurl}


\bibitem[for Cybersecurity~({ENISA})(2023)]%
        {enisaSharing}
\bibfield{author}{\bibinfo{person}{European Union~Agency for Cybersecurity~({ENISA})}.} \bibinfo{year}{2023}\natexlab{}.
\newblock \bibinfo{title}{Engineering Personal Data Sharing}.
\newblock \bibinfo{howpublished}{\url{https://www.enisa.europa.eu/publications/engineering-personal-data-sharing}}.
\newblock
\newblock
\shownote{(Accessed on 06/01/2024)}.


\bibitem[{Frederick P. Brooks, Jr.}(1975)]%
        {Brooks1975}
\bibfield{author}{\bibinfo{person}{{Frederick P. Brooks, Jr.}}} \bibinfo{year}{1975}\natexlab{}.
\newblock \bibinfo{booktitle}{\emph{The Mythical Man-Month: Essays on Software Engineering}}.
\newblock \bibinfo{publisher}{Addison-Wesley}.
\newblock


\bibitem[Frida(2022)]%
        {Frida}
\bibfield{author}{\bibinfo{person}{Frida}.} \bibinfo{year}{2022}\natexlab{}.
\newblock
\newblock
\newblock
\shownote{\url{https://frida.re/}}.


\bibitem[Gibler et~al\mbox{.}(2012)]%
        {Gibler2012}
\bibfield{author}{\bibinfo{person}{C. Gibler}, \bibinfo{person}{J. Crussell}, \bibinfo{person}{J. Erickson}, {and} \bibinfo{person}{H. Chen}.} \bibinfo{year}{2012}\natexlab{}.
\newblock \showarticletitle{{AndroidLeaks: Automatically Detecting Potential Privacy Leaks in Android Applications on a Large Scale}}. In \bibinfo{booktitle}{\emph{Proc. of the 5th international conference on Trust and Trustworthy Computing (TRUST)}}. \bibinfo{publisher}{Springer-Verlag}, \bibinfo{pages}{291--307}.
\newblock


\bibitem[GizChina(2023)]%
        {gizChinaStatement}
\bibfield{author}{\bibinfo{person}{GizChina}.} \bibinfo{year}{2023}\natexlab{}.
\newblock \bibinfo{title}{{HARMONYOS IS NOW FIRMLY THE THIRD LARGEST MOBILE PHONE OPERATING SYSTEM}}.
\newblock \bibinfo{howpublished}{\url{https://www.gizchina.com/2023/05/20/harmonyos-is-now-firmly-the-third-largest-mobile-phone-operating-system/}}.
\newblock
\newblock
\shownote{(Accessed on 01/01/2024)}.


\bibitem[Google(2023a)]%
        {baseBundle}
\bibfield{author}{\bibinfo{person}{Google}.} \bibinfo{year}{2023}\natexlab{a}.
\newblock \bibinfo{title}{BaseBundle}.
\newblock \bibinfo{howpublished}{Android Developers}.
\newblock
\newblock
\shownote{\url{https://developer.android.com/reference/android/os/BaseBundle}}.


\bibitem[Google(2023b)]%
        {GoogleSecurity}
\bibfield{author}{\bibinfo{person}{Google}.} \bibinfo{year}{2023}\natexlab{b}.
\newblock \bibinfo{title}{Design for Safety}.
\newblock \bibinfo{howpublished}{Google Developers}.
\newblock
\newblock
\shownote{\url{https://developer.android.com/quality/privacy-and-security}}.


\bibitem[Google(2023c)]%
        {firebaseMessagingService}
\bibfield{author}{\bibinfo{person}{Google}.} \bibinfo{year}{2023}\natexlab{c}.
\newblock \bibinfo{title}{FirebaseMessagingService}.
\newblock \bibinfo{howpublished}{\url{https://firebase.google.com/docs/reference/android/com/google/firebase/messaging/FirebaseMessagingService}}.
\newblock
\newblock
\shownote{(Accessed on 06/01/2024)}.


\bibitem[Google(2023d)]%
        {googlePrivacyLabels}
\bibfield{author}{\bibinfo{person}{Google}.} \bibinfo{year}{2023}\natexlab{d}.
\newblock \bibinfo{title}{{Play Console Help: Provide information for Google Play's Data safety section}}.
\newblock \bibinfo{howpublished}{\url{https://support.google.com/googleplay/android-developer/answer/10787469 }}.
\newblock
\newblock
\shownote{(Accessed on 06/01/2024)}.


\bibitem[{Google for Developers}(2024)]%
        {fcmMessage}
\bibfield{author}{\bibinfo{person}{{Google for Developers}}.} \bibinfo{year}{2024}\natexlab{}.
\newblock \bibinfo{title}{About FCM messages}.
\newblock \bibinfo{howpublished}{Developer documentation for Firebase}.
\newblock
\newblock
\shownote{\url{https://firebase.google.com/docs/cloud-messaging/concept-options}}.


\bibitem[Gordon et~al\mbox{.}(2015)]%
        {Gordon2015}
\bibfield{author}{\bibinfo{person}{M.~I. Gordon}, \bibinfo{person}{D. Kim}, \bibinfo{person}{J. Perkins}, \bibinfo{person}{Gilhamy}, \bibinfo{person}{N. Nguyenz}, {and} \bibinfo{person}{M. Rinard}.} \bibinfo{year}{2015}\natexlab{}.
\newblock \showarticletitle{{Information-Flow Analysis of Android Applications in DroidSafe}}. In \bibinfo{booktitle}{\emph{{Proc. of NDSS Symposium}}}.
\newblock


\bibitem[Hansen et~al\mbox{.}(2015)]%
        {hansen2015protection}
\bibfield{author}{\bibinfo{person}{Marit Hansen}, \bibinfo{person}{Meiko Jensen}, {and} \bibinfo{person}{Martin Rost}.} \bibinfo{year}{2015}\natexlab{}.
\newblock \showarticletitle{Protection goals for privacy engineering}. In \bibinfo{booktitle}{\emph{2015 IEEE Security and Privacy Workshops}}. IEEE, \bibinfo{pages}{159--166}.
\newblock


\bibitem[Harkous et~al\mbox{.}(2018)]%
        {harkous2018polisis}
\bibfield{author}{\bibinfo{person}{Hamza Harkous}, \bibinfo{person}{Kassem Fawaz}, \bibinfo{person}{R{\'e}mi Lebret}, \bibinfo{person}{Florian Schaub}, \bibinfo{person}{Kang~G Shin}, {and} \bibinfo{person}{Karl Aberer}.} \bibinfo{year}{2018}\natexlab{}.
\newblock \showarticletitle{Polisis: Automated analysis and presentation of privacy policies using deep learning}. In \bibinfo{booktitle}{\emph{27th USENIX Security Symposium (USENIX Security 18)}}. \bibinfo{publisher}{USENIX}, \bibinfo{address}{Berkeley, CA, USA}, \bibinfo{pages}{531--548}.
\newblock


\bibitem[{Harwell, Drew and Schaffer, Aaron}(2024)]%
        {wapoPushThreats}
\bibfield{author}{\bibinfo{person}{{Harwell, Drew and Schaffer, Aaron}}.} \bibinfo{year}{2024}\natexlab{}.
\newblock \bibinfo{title}{{The FBI’s new tactic: Catching suspects with push alerts}}.
\newblock \bibinfo{howpublished}{\url{https://www.washingtonpost.com/technology/2024/02/29/push-notification-surveillance-fbi/}}.
\newblock
\newblock
\shownote{(Accessed on 06/01/2024)}.


\bibitem[Hyun et~al\mbox{.}(2018)]%
        {hyun2018design}
\bibfield{author}{\bibinfo{person}{Sangwon Hyun}, \bibinfo{person}{Junsung Cho}, \bibinfo{person}{Geumhwan Cho}, {and} \bibinfo{person}{Hyoungshick Kim}.} \bibinfo{year}{2018}\natexlab{}.
\newblock \showarticletitle{Design and analysis of push notification-based malware on android}.
\newblock \bibinfo{journal}{\emph{Security and Communication Networks}}  \bibinfo{volume}{2018} (\bibinfo{year}{2018}).
\newblock


\bibitem[JusTalk(2023)]%
        {jusTalkStatement}
\bibfield{author}{\bibinfo{person}{JusTalk}.} \bibinfo{year}{2023}\natexlab{}.
\newblock \bibinfo{title}{{Is it safe to use JusTalk?}}
\newblock \bibinfo{howpublished}{\url{https://web.archive.org/web/20230407183707/https://justalk.com/support/general/g6}}.
\newblock
\newblock
\shownote{(Accessed on 10/10/2023)}.


\bibitem[Kelley et~al\mbox{.}(2013)]%
        {kelley2013privacy}
\bibfield{author}{\bibinfo{person}{P.~G. Kelley}, \bibinfo{person}{L.~F. Cranor}, {and} \bibinfo{person}{N. Sadeh}.} \bibinfo{year}{2013}\natexlab{}.
\newblock \showarticletitle{Privacy as part of the app decision-making process}. In \bibinfo{booktitle}{\emph{Proceedings of the SIGCHI conference on human factors in computing systems}}. \bibinfo{pages}{3393--3402}.
\newblock


\bibitem[Kim et~al\mbox{.}(2012)]%
        {kim2012scandal}
\bibfield{author}{\bibinfo{person}{J. Kim}, \bibinfo{person}{Y. Yoon}, \bibinfo{person}{K. Yi}, {and} \bibinfo{person}{J. Shin}.} \bibinfo{year}{2012}\natexlab{}.
\newblock \showarticletitle{{ScanDal: Static Analyzer for Detecting Privacy Leaks in Android Applications}}.
\newblock \bibinfo{journal}{\emph{{IEEE Workshop on Mobile Security Technologies (MoST)}}} (\bibinfo{year}{2012}).
\newblock


\bibitem[Koch et~al\mbox{.}(2022)]%
        {koch2022keeping}
\bibfield{author}{\bibinfo{person}{Simon Koch}, \bibinfo{person}{Malte Wessels}, \bibinfo{person}{Benjamin Altpeter}, \bibinfo{person}{Madita Olvermann}, {and} \bibinfo{person}{Martin Johns}.} \bibinfo{year}{2022}\natexlab{}.
\newblock \showarticletitle{Keeping privacy labels honest}.
\newblock \bibinfo{journal}{\emph{Proceedings on Privacy Enhancing Technologies}} \bibinfo{volume}{4}, \bibinfo{number}{486-506} (\bibinfo{year}{2022}), \bibinfo{pages}{2--2}.
\newblock


\bibitem[{Konev, Max}(2022)]%
        {pushwooshStatement}
\bibfield{author}{\bibinfo{person}{{Konev, Max}}.} \bibinfo{year}{2022}\natexlab{}.
\newblock \bibinfo{title}{{Statement on the Reuters Story Regarding Pushwoosh}}.
\newblock \bibinfo{howpublished}{\url{https://blog.pushwoosh.com/blog/statement-on-the-reuters-story-regarding-pushwoosh/}}.
\newblock
\newblock
\shownote{(Accessed on 06/01/2024)}.


\bibitem[Lee et~al\mbox{.}(2014)]%
        {lee2014punobot}
\bibfield{author}{\bibinfo{person}{Hayoung Lee}, \bibinfo{person}{Taeho Kang}, \bibinfo{person}{Sangho Lee}, \bibinfo{person}{Jong Kim}, {and} \bibinfo{person}{Yoonho Kim}.} \bibinfo{year}{2014}\natexlab{}.
\newblock \showarticletitle{Punobot: Mobile botnet using push notification service in android}. In \bibinfo{booktitle}{\emph{Information Security Applications: 14th International Workshop, WISA 2013, Jeju Island, Korea, August 19-21, 2013, Revised Selected Papers 14}}. Springer, \bibinfo{pages}{124--137}.
\newblock


\bibitem[Li et~al\mbox{.}(2014)]%
        {li2014mayhem}
\bibfield{author}{\bibinfo{person}{Tongxin Li}, \bibinfo{person}{Xiaoyong Zhou}, \bibinfo{person}{Luyi Xing}, \bibinfo{person}{Yeonjoon Lee}, \bibinfo{person}{Muhammad Naveed}, \bibinfo{person}{XiaoFeng Wang}, {and} \bibinfo{person}{Xinhui Han}.} \bibinfo{year}{2014}\natexlab{}.
\newblock \showarticletitle{Mayhem in the push clouds: Understanding and mitigating security hazards in mobile push-messaging services}. In \bibinfo{booktitle}{\emph{Proceedings of the 2014 ACM SIGSAC Conference on Computer and Communications Security}}. \bibinfo{pages}{978--989}.
\newblock


\bibitem[Linden et~al\mbox{.}(2018)]%
        {linden2018privacy}
\bibfield{author}{\bibinfo{person}{Thomas Linden}, \bibinfo{person}{Rishabh Khandelwal}, \bibinfo{person}{Hamza Harkous}, {and} \bibinfo{person}{Kassem Fawaz}.} \bibinfo{year}{2018}\natexlab{}.
\newblock \showarticletitle{The privacy policy landscape after the GDPR}.
\newblock \bibinfo{journal}{\emph{arXiv preprint arXiv:1809.08396}} (\bibinfo{year}{2018}), \bibinfo{pages}{1--18}.
\newblock


\bibitem[Liu et~al\mbox{.}(2019)]%
        {liu2019dapanda}
\bibfield{author}{\bibinfo{person}{Tianming Liu}, \bibinfo{person}{Haoyu Wang}, \bibinfo{person}{Li Li}, \bibinfo{person}{Guangdong Bai}, \bibinfo{person}{Yao Guo}, {and} \bibinfo{person}{Guoai Xu}.} \bibinfo{year}{2019}\natexlab{}.
\newblock \showarticletitle{Dapanda: Detecting aggressive push notifications in android apps}. In \bibinfo{booktitle}{\emph{2019 34th IEEE/ACM International Conference on Automated Software Engineering (ASE)}}. IEEE, \bibinfo{pages}{66--78}.
\newblock


\bibitem[Loreti et~al\mbox{.}(2018)]%
        {loreti2018push}
\bibfield{author}{\bibinfo{person}{Pierpaolo Loreti}, \bibinfo{person}{Lorenzo Bracciale}, {and} \bibinfo{person}{Alberto Caponi}.} \bibinfo{year}{2018}\natexlab{}.
\newblock \showarticletitle{Push attack: binding virtual and real identities using mobile push notifications}.
\newblock \bibinfo{journal}{\emph{Future Internet}} \bibinfo{volume}{10}, \bibinfo{number}{2} (\bibinfo{year}{2018}), \bibinfo{pages}{13}.
\newblock


\bibitem[Lou et~al\mbox{.}(2023)]%
        {lou2023devils}
\bibfield{author}{\bibinfo{person}{Jiadong Lou}, \bibinfo{person}{Xiaohan Zhang}, \bibinfo{person}{Yihe Zhang}, \bibinfo{person}{Xinghua Li}, \bibinfo{person}{Xu Yuan}, {and} \bibinfo{person}{Ning Zhang}.} \bibinfo{year}{2023}\natexlab{}.
\newblock \showarticletitle{Devils in Your Apps: Vulnerabilities and User Privacy Exposure in Mobile Notification Systems}. In \bibinfo{booktitle}{\emph{2023 53rd Annual IEEE/IFIP International Conference on Dependable Systems and Networks (DSN)}}. IEEE, \bibinfo{pages}{28--41}.
\newblock


\bibitem[Madden(2014)]%
        {Madden2014}
\bibfield{author}{\bibinfo{person}{Mary Madden}.} \bibinfo{year}{2014}\natexlab{}.
\newblock \bibinfo{title}{Public Perceptions of Privacy and Security in the Post-Snowden Era}.
\newblock \bibinfo{howpublished}{Pew Research Center}.
\newblock
\newblock
\shownote{\url{https://www.pewresearch.org/internet/2014/11/12/public-privacy-perceptions/}}.


\bibitem[Marx(1875)]%
        {Marx1875}
\bibfield{author}{\bibinfo{person}{Karl Marx}.} \bibinfo{year}{1875}\natexlab{}.
\newblock \bibinfo{booktitle}{\emph{Critique of the Gotha program}}.
\newblock


\bibitem[McLuhan(1964)]%
        {McLuhan1964}
\bibfield{author}{\bibinfo{person}{Marshall McLuhan}.} \bibinfo{year}{1964}\natexlab{}.
\newblock \showarticletitle{Understanding Media}.
\newblock  (\bibinfo{year}{1964}).
\newblock


\bibitem[Microsoft(2023)]%
        {skypeStatement}
\bibfield{author}{\bibinfo{person}{Microsoft}.} \bibinfo{year}{2023}\natexlab{}.
\newblock \bibinfo{title}{{What are Skype Private Conversations?}}
\newblock \bibinfo{howpublished}{\url{https://web.archive.org/web/20230606085952/https://support.skype.com/en/faq/fa34824/what-are-skype-private-conversations}}.
\newblock
\newblock
\shownote{(Accessed on 10/10/2023)}.


\bibitem[Okoyomon et~al\mbox{.}(2019)]%
        {okoyomon2019ridiculousness}
\bibfield{author}{\bibinfo{person}{Ehimare Okoyomon}, \bibinfo{person}{Nikita Samarin}, \bibinfo{person}{Primal Wijesekera}, \bibinfo{person}{Amit Elazari Bar~On}, \bibinfo{person}{Narseo Vallina-Rodriguez}, \bibinfo{person}{Irwin Reyes}, \bibinfo{person}{{\'A}lvaro Feal}, \bibinfo{person}{Serge Egelman}, {et~al\mbox{.}}} \bibinfo{year}{2019}\natexlab{}.
\newblock \showarticletitle{On the ridiculousness of notice and consent: Contradictions in app privacy policies}. In \bibinfo{booktitle}{\emph{Workshop on Technology and Consumer Protection (ConPro 2019), in conjunction with the 39th IEEE Symposium on Security and Privacy}}. \bibinfo{publisher}{IEEE}, \bibinfo{address}{New York, NY, USA}.
\newblock


\bibitem[OneSignal(2023a)]%
        {oneSignalStatement}
\bibfield{author}{\bibinfo{person}{OneSignal}.} \bibinfo{year}{2023}\natexlab{a}.
\newblock \bibinfo{title}{{Firebase Cloud Messaging (FCM) Compared to OneSignal}}.
\newblock \bibinfo{howpublished}{\url{https://web.archive.org/web/20230603040346/https://onesignal.com/blog/firebase-vs-onesignal/}}.
\newblock
\newblock
\shownote{(Accessed on 10/10/2023)}.


\bibitem[OneSignal(2023b)]%
        {oneSignalPush}
\bibfield{author}{\bibinfo{person}{OneSignal}.} \bibinfo{year}{2023}\natexlab{b}.
\newblock \bibinfo{title}{What is a push notifications service and how does it work?}
\newblock \bibinfo{howpublished}{\url{https://onesignal.com/blog/what-is-a-push-notifications-service-and-how-does-it-work/}}.
\newblock
\newblock
\shownote{(Accessed on 2/23/24)}.


\bibitem[Pallas et~al\mbox{.}(2024)]%
        {pallas2024privacy}
\bibfield{author}{\bibinfo{person}{Frank Pallas}, \bibinfo{person}{Katharina Koerner}, \bibinfo{person}{Isabel Barber{\'a}}, \bibinfo{person}{Jaap-Henk Hoepman}, \bibinfo{person}{Meiko Jensen}, \bibinfo{person}{Nandita~Rao Narla}, \bibinfo{person}{Nikita Samarin}, \bibinfo{person}{Max-R Ulbricht}, \bibinfo{person}{Isabel Wagner}, \bibinfo{person}{Kim Wuyts}, {et~al\mbox{.}}} \bibinfo{year}{2024}\natexlab{}.
\newblock \showarticletitle{Privacy Engineering From Principles to Practice: A Roadmap}.
\newblock \bibinfo{journal}{\emph{IEEE Security \& Privacy}} \bibinfo{volume}{22}, \bibinfo{number}{2} (\bibinfo{year}{2024}), \bibinfo{pages}{86--92}.
\newblock


\bibitem[Pearson and Taylor(2022)]%
        {reutersPushwoosh}
\bibfield{author}{\bibinfo{person}{James Pearson} {and} \bibinfo{person}{Marisa Taylor}.} \bibinfo{year}{2022}\natexlab{}.
\newblock \bibinfo{title}{{Russian software disguised as American finds its way into U.S. Army, CDC apps}}.
\newblock \bibinfo{howpublished}{\url{https://www.reuters.com/technology/exclusive-russian-software-disguised-american-finds-its-way-into-us-army-cdc-2022-11-14/}}.
\newblock
\newblock
\shownote{(Accessed on 06/01/2024)}.


\bibitem[Play(2023)]%
        {weChatDescription}
\bibfield{author}{\bibinfo{person}{Google Play}.} \bibinfo{year}{2023}\natexlab{}.
\newblock \bibinfo{title}{{WeChat: About this app}}.
\newblock \bibinfo{howpublished}{\url{https://web.archive.org/web/20230323082225/https://play.google.com/store/apps/details?id=com.tencent.mm&hl=en_US&gl=US}}.
\newblock
\newblock
\shownote{(Accessed on 10/10/2023)}.


\bibitem[Pusher(2023)]%
        {pusherStatement}
\bibfield{author}{\bibinfo{person}{Pusher}.} \bibinfo{year}{2023}\natexlab{}.
\newblock \bibinfo{title}{{Configure FCM}}.
\newblock \bibinfo{howpublished}{\url{https://pusher.com/docs/beams/getting-started/android/configure-fcm/}}.
\newblock
\newblock
\shownote{(Accessed on 10/10/2023)}.


\bibitem[Razaghpanah et~al\mbox{.}(2017)]%
        {razaghpanah2017studying}
\bibfield{author}{\bibinfo{person}{A. Razaghpanah}, \bibinfo{person}{A.~A. Niaki}, \bibinfo{person}{N. Vallina-Rodriguez}, \bibinfo{person}{S. Sundaresan}, \bibinfo{person}{J. Amann}, {and} \bibinfo{person}{P. Gill}.} \bibinfo{year}{2017}\natexlab{}.
\newblock \showarticletitle{{Studying TLS usage in Android apps}}. In \bibinfo{booktitle}{\emph{Proceedings of the 13th International Conference on emerging Networking EXperiments and Technologies}}. \bibinfo{pages}{350--362}.
\newblock


\bibitem[Reyes et~al\mbox{.}(2018)]%
        {reyes2018won}
\bibfield{author}{\bibinfo{person}{Irwin Reyes}, \bibinfo{person}{Primal Wijesekera}, \bibinfo{person}{Joel Reardon}, \bibinfo{person}{Amit Elazari Bar~On}, \bibinfo{person}{Abbas Razaghpanah}, \bibinfo{person}{Narseo Vallina-Rodriguez}, \bibinfo{person}{Serge Egelman}, {et~al\mbox{.}}} \bibinfo{year}{2018}\natexlab{}.
\newblock \showarticletitle{``Won't somebody think of the children?'' examining COPPA compliance at scale}.
\newblock \bibinfo{journal}{\emph{Proceedings on Privacy Enhancing Technologies (PoPETs)}} \bibinfo{volume}{2018}, \bibinfo{number}{3} (\bibinfo{year}{2018}), \bibinfo{pages}{63--83}.
\newblock


\bibitem[Rodriguez et~al\mbox{.}(2023)]%
        {rodriguez2023comparing}
\bibfield{author}{\bibinfo{person}{David Rodriguez}, \bibinfo{person}{Akshath Jain}, \bibinfo{person}{Jose~M Del~Alamo}, {and} \bibinfo{person}{Norman Sadeh}.} \bibinfo{year}{2023}\natexlab{}.
\newblock \showarticletitle{Comparing Privacy Label Disclosures of Apps Published in both the App Store and Google Play Stores}. In \bibinfo{booktitle}{\emph{2023 IEEE European Symposium on Security and Privacy Workshops (EuroS\&PW)}}. IEEE, \bibinfo{pages}{150--157}.
\newblock


\bibitem[SafeUM(2023)]%
        {safeumPolicy}
\bibfield{author}{\bibinfo{person}{SafeUM}.} \bibinfo{year}{2023}\natexlab{}.
\newblock \bibinfo{title}{{Privacy Policy}}.
\newblock \bibinfo{howpublished}{\url{https://web.archive.org/web/20230220213832/https://safeum.com/privacypolicy.html}}.
\newblock
\newblock
\shownote{(Accessed on 10/10/2023)}.


\bibitem[Samarin et~al\mbox{.}(2023)]%
        {samarin2023lessons}
\bibfield{author}{\bibinfo{person}{Nikita Samarin}, \bibinfo{person}{Shayna Kothari}, \bibinfo{person}{Zaina Siyed}, \bibinfo{person}{Oscar Bjorkman}, \bibinfo{person}{Reena Yuan}, \bibinfo{person}{Primal Wijesekera}, \bibinfo{person}{Noura Alomar}, \bibinfo{person}{Jordan Fischer}, \bibinfo{person}{Chris Hoofnagle}, {and} \bibinfo{person}{Serge Egelman}.} \bibinfo{year}{2023}\natexlab{}.
\newblock \showarticletitle{Lessons in VCR Repair: Compliance of Android App Developers with the California Consumer Privacy Act (CCPA)}.
\newblock \bibinfo{journal}{\emph{arXiv preprint arXiv:2304.00944}} (\bibinfo{year}{2023}).
\newblock


\bibitem[Shi(2023)]%
        {androidDevBlog}
\bibfield{author}{\bibinfo{person}{Jingyu Shi}.} \bibinfo{year}{2023}\natexlab{}.
\newblock \bibinfo{title}{{Notifying your users with FCM}}.
\newblock \bibinfo{howpublished}{\url{https://android-developers.googleblog.com/2018/09/notifying-your-users-with-fcm.html}}.
\newblock
\newblock
\shownote{(Accessed on 10/10/2023)}.


\bibitem[Signal(2023a)]%
        {signalGrandJury}
\bibfield{author}{\bibinfo{person}{Signal}.} \bibinfo{year}{2023}\natexlab{a}.
\newblock \bibinfo{title}{{Grand jury subpoena for Signal user data, Central District of California (again!)}}.
\newblock \bibinfo{howpublished}{\url{https://web.archive.org/web/20230921202338/https://signal.org/bigbrother/cd-california-grand-jury/}}.
\newblock
\newblock
\shownote{(Accessed on 10/10/2023)}.


\bibitem[Signal(2023b)]%
        {signalMainPage}
\bibfield{author}{\bibinfo{person}{Signal}.} \bibinfo{year}{2023}\natexlab{b}.
\newblock \bibinfo{title}{{Signal}}.
\newblock \bibinfo{howpublished}{\url{https://signal.org/}}.
\newblock
\newblock
\shownote{(Accessed on 10/10/2023)}.


\bibitem[Slavin et~al\mbox{.}(2016)]%
        {slavin2016toward}
\bibfield{author}{\bibinfo{person}{Rocky Slavin}, \bibinfo{person}{Xiaoyin Wang}, \bibinfo{person}{Mitra~Bokaei Hosseini}, \bibinfo{person}{James Hester}, \bibinfo{person}{Ram Krishnan}, \bibinfo{person}{Jaspreet Bhatia}, \bibinfo{person}{Travis~D Breaux}, {and} \bibinfo{person}{Jianwei Niu}.} \bibinfo{year}{2016}\natexlab{}.
\newblock \showarticletitle{Toward a framework for detecting privacy policy violations in android application code}. In \bibinfo{booktitle}{\emph{Proceedings of the 38th International Conference on Software Engineering}}. \bibinfo{publisher}{ACM}, \bibinfo{address}{New York, NY, USA}, \bibinfo{pages}{25--36}.
\newblock


\bibitem[Slobozhan et~al\mbox{.}(2023)]%
        {slobozhan2023differentiable}
\bibfield{author}{\bibinfo{person}{Ivan Slobozhan}, \bibinfo{person}{Tymofii Brik}, {and} \bibinfo{person}{Rajesh Sharma}.} \bibinfo{year}{2023}\natexlab{}.
\newblock \showarticletitle{Differentiable characteristics of Telegram mediums during protests in Belarus 2020}.
\newblock \bibinfo{journal}{\emph{Social Network Analysis and Mining}} \bibinfo{volume}{13}, \bibinfo{number}{1} (\bibinfo{year}{2023}), \bibinfo{pages}{19}.
\newblock


\bibitem[Smith(1776)]%
        {smith1776inquiry}
\bibfield{author}{\bibinfo{person}{Adam Smith}.} \bibinfo{year}{1776}\natexlab{}.
\newblock \bibinfo{booktitle}{\emph{An Inquiry Into the Nature and Causes of the Wealth of Nations}}.
\newblock \bibinfo{publisher}{Strahan and Cadell}, \bibinfo{address}{London, UK}.
\newblock
\showISBNx{9781546508649}
\urldef\tempurl%
\url{https://books.google.com/books?id=mt1SAAAAcAAJ}
\showURL{%
\tempurl}


\bibitem[{StatCounter Global Stats}(2023)]%
        {androidMarketShare}
\bibfield{author}{\bibinfo{person}{{StatCounter Global Stats}}.} \bibinfo{year}{2023}\natexlab{}.
\newblock \bibinfo{title}{{Android Version Market Share Worldwide}}.
\newblock \bibinfo{howpublished}{\url{https://gs.statcounter.com/android-version-market-share/all/worldwide/2023}}.
\newblock
\newblock
\shownote{(Accessed on 06/01/2024)}.


\bibitem[Stopper and Caltrider(2023)]%
        {stoppersee}
\bibfield{author}{\bibinfo{person}{Anne Stopper} {and} \bibinfo{person}{Jen Caltrider}.} \bibinfo{year}{2023}\natexlab{}.
\newblock \bibinfo{title}{See no evil: Loopholes in Google’s data safety labels keep companies in the clear and consumers in the dark. mozilla foundation}.
\newblock
\newblock


\bibitem[Tan et~al\mbox{.}(2014)]%
        {Tan2014}
\bibfield{author}{\bibinfo{person}{J. Tan}, \bibinfo{person}{K. Nguyen}, \bibinfo{person}{M. Theodorides}, \bibinfo{person}{H. Negron-Arroyo}, \bibinfo{person}{C. Thompson}, \bibinfo{person}{S. Egelman}, {and} \bibinfo{person}{D. Wagner}.} \bibinfo{year}{2014}\natexlab{}.
\newblock \showarticletitle{{The Effect of Developer-Specified Explanations for Permission Requests on Smartphone User Behavior}}. In \bibinfo{booktitle}{\emph{Proceedings of the SIGCHI Conference on Human Factors in Computing Systems}}.
\newblock


\bibitem[Telegram(2023)]%
        {telegramMainPage}
\bibfield{author}{\bibinfo{person}{Telegram}.} \bibinfo{year}{2023}\natexlab{}.
\newblock \bibinfo{title}{{Telegram Messenger}}.
\newblock \bibinfo{howpublished}{\url{https://telegram.org/}}.
\newblock
\newblock
\shownote{(Accessed on 10/10/2023)}.


\bibitem[{Telegram-FOSS on GitHub}(2024)]%
        {telegramPersistentPush}
\bibfield{author}{\bibinfo{person}{{Telegram-FOSS on GitHub}}.} \bibinfo{year}{2024}\natexlab{}.
\newblock \bibinfo{title}{{Notifications}}.
\newblock \bibinfo{howpublished}{\url{https://github.com/Telegram-FOSS-Team/Telegram-FOSS/blob/master/Notifications.md}}.
\newblock
\newblock
\shownote{(Accessed on 06/01/2024)}.


\bibitem[{The Drum}(2023)]%
        {whatsappBillboard}
\bibfield{author}{\bibinfo{person}{{The Drum}}.} \bibinfo{year}{2023}\natexlab{}.
\newblock \bibinfo{title}{{WhatsApp’s 3D billboard touts privacy features}}.
\newblock \bibinfo{howpublished}{\url{https://www.thedrum.com/news/2022/10/10/whatsapp-s-3d-billboard-touts-privacy-features}}.
\newblock
\newblock
\shownote{(Accessed on 10/10/2023)}.


\bibitem[{The Verge}(2023)]%
        {whatsappFunApple}
\bibfield{author}{\bibinfo{person}{{The Verge}}.} \bibinfo{year}{2023}\natexlab{}.
\newblock \bibinfo{title}{{Now Mark Zuckerberg’s making fun of Apple for iMessage, too}}.
\newblock \bibinfo{howpublished}{\url{https://www.theverge.com/2022/10/17/23409018/mark-zuckerberg-meta-whatsapp-imessage-privacy-security-ads}}.
\newblock
\newblock
\shownote{(Accessed on 10/10/2023)}.


\bibitem[{The White House}(2023)]%
        {whiteHouseStrategy}
\bibfield{author}{\bibinfo{person}{{The White House}}.} \bibinfo{year}{2023}\natexlab{}.
\newblock \bibinfo{title}{{National Cybersecurity Strategy}}.
\newblock \bibinfo{howpublished}{\url{https://www.whitehouse.gov/wp-content/uploads/2023/03/National-Cybersecurity-Strategy-2023.pdf}}.
\newblock
\newblock
\shownote{(Accessed on 06/01/2024)}.


\bibitem[Thompson et~al\mbox{.}(2013)]%
        {Thompson2013}
\bibfield{author}{\bibinfo{person}{C. Thompson}, \bibinfo{person}{M. Johnson}, \bibinfo{person}{S. Egelman}, \bibinfo{person}{D. Wagner}, {and} \bibinfo{person}{J. King}.} \bibinfo{year}{2013}\natexlab{}.
\newblock \showarticletitle{{When It's Better to Ask Forgiveness than Get Permission: Designing Usable Audit Mechanisms for Mobile Permissions}}. In \bibinfo{booktitle}{\emph{Proceedings of the 2013 Symposium on Usable Privacy and Security (SOUPS)}}.
\newblock


\bibitem[Tsai et~al\mbox{.}(2017)]%
        {Tsai2017}
\bibfield{author}{\bibinfo{person}{L. Tsai}, \bibinfo{person}{P. Wijesekera}, \bibinfo{person}{J. Reardon}, \bibinfo{person}{I. Reyes}, \bibinfo{person}{S. Egelman}, \bibinfo{person}{D. Wagner}, \bibinfo{person}{N. Good}, {and} \bibinfo{person}{J. Chen}.} \bibinfo{year}{2017}\natexlab{}.
\newblock \showarticletitle{{Turtle Guard: Helping Android Users Apply Contextual Privacy Preferences}}. In \bibinfo{booktitle}{\emph{Thirteenth Symposium on Usable Privacy and Security ({SOUPS} 2017)}}. \bibinfo{publisher}{{USENIX} Association}, \bibinfo{address}{Santa Clara, CA}, \bibinfo{pages}{145--162}.
\newblock
\showISBNx{978-1-931971-39-3}
\urldef\tempurl%
\url{https://www.usenix.org/conference/soups2017/technical-sessions/presentation/tsai}
\showURL{%
\tempurl}


\bibitem[Tufekci(2017)]%
        {tufekci2017twitter}
\bibfield{author}{\bibinfo{person}{Zeynep Tufekci}.} \bibinfo{year}{2017}\natexlab{}.
\newblock \bibinfo{booktitle}{\emph{Twitter and tear gas: The power and fragility of networked protest}}.
\newblock \bibinfo{publisher}{Yale University Press}.
\newblock


\bibitem[{UnifiedPush}(2023)]%
        {unifiedPush}
\bibfield{author}{\bibinfo{person}{{UnifiedPush}}.} \bibinfo{year}{2023}\natexlab{}.
\newblock \bibinfo{title}{{UnifiedPush}}.
\newblock \bibinfo{howpublished}{\url{https://unifiedpush.org/}}.
\newblock
\newblock
\shownote{(Accessed on 10/10/2023)}.


\bibitem[{United States District Court for the Central District of California}(2022)]%
        {pushWarrant1}
\bibfield{author}{\bibinfo{person}{{United States District Court for the Central District of California}}.} \bibinfo{year}{2022}\natexlab{}.
\newblock \bibinfo{title}{{Application for a Warrant re: Case No. 2:22-MJ-03119}}.
\newblock \bibinfo{howpublished}{\url{https://www.documentcloud.org/documents/24192891-search-warrant-for-google-account-for-push-notification-data}}.
\newblock
\newblock
\shownote{(Accessed on 06/01/2024)}.


\bibitem[{United States District Court for the District of Columbia}(2021)]%
        {pushWarrant2}
\bibfield{author}{\bibinfo{person}{{United States District Court for the District of Columbia}}.} \bibinfo{year}{2021}\natexlab{}.
\newblock \bibinfo{title}{{Application for a Warrant re: Case No. 21-sc-270}}.
\newblock \bibinfo{howpublished}{\url{https://www.documentcloud.org/documents/24192911-6d68977d-f8ef-4080-9742-290cff8a6c28}}.
\newblock
\newblock
\shownote{(Accessed on 06/01/2024)}.


\bibitem[Urman et~al\mbox{.}(2021)]%
        {urman2021analyzing}
\bibfield{author}{\bibinfo{person}{Aleksandra Urman}, \bibinfo{person}{Justin Chun-ting Ho}, {and} \bibinfo{person}{Stefan Katz}.} \bibinfo{year}{2021}\natexlab{}.
\newblock \showarticletitle{Analyzing protest mobilization on Telegram: The case of 2019 anti-extradition bill movement in Hong Kong}.
\newblock \bibinfo{journal}{\emph{Plos one}} \bibinfo{volume}{16}, \bibinfo{number}{10} (\bibinfo{year}{2021}), \bibinfo{pages}{e0256675}.
\newblock


\bibitem[{U.S. Congress}(1986)]%
        {ecpa}
\bibfield{author}{\bibinfo{person}{{U.S. Congress}}.} \bibinfo{year}{1986}\natexlab{}.
\newblock \bibinfo{title}{{H.R.4952 - Electronic Communications Privacy Act of 1986 }}.
\newblock \bibinfo{howpublished}{\url{https://www.congress.gov/bill/99th-congress/house-bill/4952}}.
\newblock
\newblock
\shownote{(Accessed on 10/10/2023)}.


\bibitem[Viber(2023)]%
        {viberPolicy}
\bibfield{author}{\bibinfo{person}{Viber}.} \bibinfo{year}{2023}\natexlab{}.
\newblock \bibinfo{title}{{Privacy Notice for California Residents}}.
\newblock \bibinfo{howpublished}{\url{https://web.archive.org/web/20230310001732/https://www.viber.com/en/terms/ccpa-privacy-rights/}}.
\newblock
\newblock
\shownote{(Accessed on 10/10/2023)}.


\bibitem[Wang et~al\mbox{.}(2018)]%
        {wang2018guileak}
\bibfield{author}{\bibinfo{person}{Xiaoyin Wang}, \bibinfo{person}{Xue Qin}, \bibinfo{person}{Mitra~Bokaei Hosseini}, \bibinfo{person}{Rocky Slavin}, \bibinfo{person}{Travis~D Breaux}, {and} \bibinfo{person}{Jianwei Niu}.} \bibinfo{year}{2018}\natexlab{}.
\newblock \showarticletitle{Guileak: Tracing privacy policy claims on user input data for android applications}. In \bibinfo{booktitle}{\emph{Proceedings of the 40th International Conference on Software Engineering}}. \bibinfo{publisher}{ACM}, \bibinfo{address}{New York, NY, USA}, \bibinfo{pages}{37--47}.
\newblock


\bibitem[Warren et~al\mbox{.}(2014)]%
        {warren2014push}
\bibfield{author}{\bibinfo{person}{Ian Warren}, \bibinfo{person}{Andrew Meads}, \bibinfo{person}{Satish Srirama}, \bibinfo{person}{Thiranjith Weerasinghe}, {and} \bibinfo{person}{Carlos Paniagua}.} \bibinfo{year}{2014}\natexlab{}.
\newblock \showarticletitle{Push notification mechanisms for pervasive smartphone applications}.
\newblock \bibinfo{journal}{\emph{IEEE Pervasive Computing}} \bibinfo{volume}{13}, \bibinfo{number}{2} (\bibinfo{year}{2014}), \bibinfo{pages}{61--71}.
\newblock


\bibitem[Wickham(2018)]%
        {wickham2018push}
\bibfield{author}{\bibinfo{person}{Mark Wickham}.} \bibinfo{year}{2018}\natexlab{}.
\newblock \showarticletitle{Push Messaging}.
\newblock \bibinfo{journal}{\emph{Practical Android: 14 Complete Projects on Advanced Techniques and Approaches}} (\bibinfo{year}{2018}), \bibinfo{pages}{135--172}.
\newblock


\bibitem[Wijesekera et~al\mbox{.}(2015)]%
        {wijesekera2015android}
\bibfield{author}{\bibinfo{person}{Primal Wijesekera}, \bibinfo{person}{Arjun Baokar}, \bibinfo{person}{Ashkan Hosseini}, \bibinfo{person}{Serge Egelman}, \bibinfo{person}{David Wagner}, {and} \bibinfo{person}{Konstantin Beznosov}.} \bibinfo{year}{2015}\natexlab{}.
\newblock \showarticletitle{Android permissions remystified: A field study on contextual integrity}. In \bibinfo{booktitle}{\emph{24th USENIX Security Symposium (USENIX Security 15)}}. \bibinfo{pages}{499--514}.
\newblock


\bibitem[Wijesekera et~al\mbox{.}(2017)]%
        {wijesekera2017feasibility}
\bibfield{author}{\bibinfo{person}{Primal Wijesekera}, \bibinfo{person}{Arjun Baokar}, \bibinfo{person}{Lynn Tsai}, \bibinfo{person}{Joel Reardon}, \bibinfo{person}{Serge Egelman}, \bibinfo{person}{David Wagner}, {and} \bibinfo{person}{Konstantin Beznosov}.} \bibinfo{year}{2017}\natexlab{}.
\newblock \showarticletitle{The feasibility of dynamically granted permissions: Aligning mobile privacy with user preferences}. In \bibinfo{booktitle}{\emph{2017 IEEE Symposium on Security and Privacy (SP)}}. \bibinfo{publisher}{IEEE}, \bibinfo{address}{New York, NY, USA}, \bibinfo{pages}{1077--1093}.
\newblock


\bibitem[Wijesekera et~al\mbox{.}(2018)]%
        {wijesekera2018contextualizing}
\bibfield{author}{\bibinfo{person}{Primal Wijesekera}, \bibinfo{person}{Joel Reardon}, \bibinfo{person}{Irwin Reyes}, \bibinfo{person}{Lynn Tsai}, \bibinfo{person}{Jung-Wei Chen}, \bibinfo{person}{Nathan Good}, \bibinfo{person}{David Wagner}, \bibinfo{person}{Konstantin Beznosov}, {and} \bibinfo{person}{Serge Egelman}.} \bibinfo{year}{2018}\natexlab{}.
\newblock \showarticletitle{Contextualizing privacy decisions for better prediction (and protection)}. In \bibinfo{booktitle}{\emph{Proceedings of the 2018 CHI Conference on Human Factors in Computing Systems}}. \bibinfo{pages}{1--13}.
\newblock


\bibitem[Wikipedia(2023)]%
        {ChickenGun}
\bibfield{author}{\bibinfo{person}{Wikipedia}.} \bibinfo{year}{2023}\natexlab{}.
\newblock \bibinfo{title}{Chicken Gun}.
\newblock
\newblock
\newblock
\shownote{\url{https://en.wikipedia.org/wiki/Chicken_gun}}.


\bibitem[Wuyts et~al\mbox{.}(2020)]%
        {wuyts2020linddun}
\bibfield{author}{\bibinfo{person}{Kim Wuyts}, \bibinfo{person}{Laurens Sion}, {and} \bibinfo{person}{Wouter Joosen}.} \bibinfo{year}{2020}\natexlab{}.
\newblock \showarticletitle{Linddun go: A lightweight approach to privacy threat modeling}. In \bibinfo{booktitle}{\emph{2020 IEEE European Symposium on Security and Privacy Workshops (EuroS\&PW)}}. IEEE, \bibinfo{pages}{302--309}.
\newblock


\bibitem[Wyden(2023)]%
        {wydenLetter}
\bibfield{author}{\bibinfo{person}{Ron Wyden}.} \bibinfo{year}{2023}\natexlab{}.
\newblock \bibinfo{title}{{Wyden Smartphone Push Notification Surveillance Letter}}.
\newblock \bibinfo{howpublished}{\url{https://www.wyden.senate.gov/imo/media/doc/wyden_smartphone_push_notification_surveillance_letter.pdf}}.
\newblock
\newblock
\shownote{(Accessed on 01/01/2024)}.


\bibitem[Xiao et~al\mbox{.}(2022)]%
        {xiao2022lalaine}
\bibfield{author}{\bibinfo{person}{Yue Xiao}, \bibinfo{person}{Zhengyi Li}, \bibinfo{person}{Yue Qin}, \bibinfo{person}{Xiaolong Bai}, \bibinfo{person}{Jiale Guan}, \bibinfo{person}{Xiaojing Liao}, {and} \bibinfo{person}{Luyi Xing}.} \bibinfo{year}{2022}\natexlab{}.
\newblock \showarticletitle{Lalaine: Measuring and characterizing non-compliance of apple privacy labels at scale}.
\newblock \bibinfo{journal}{\emph{arXiv preprint arXiv:2206.06274}} (\bibinfo{year}{2022}).
\newblock


\bibitem[Xu and Zhu(2012)]%
        {xu2012abusing}
\bibfield{author}{\bibinfo{person}{Zhi Xu} {and} \bibinfo{person}{Sencun Zhu}.} \bibinfo{year}{2012}\natexlab{}.
\newblock \showarticletitle{Abusing Notification Services on Smartphones for Phishing and Spamming.}. In \bibinfo{booktitle}{\emph{WOOT}}. \bibinfo{pages}{1--11}.
\newblock


\bibitem[Zimmeck et~al\mbox{.}(2021)]%
        {zimmeck2021privacyflash}
\bibfield{author}{\bibinfo{person}{Sebastian Zimmeck}, \bibinfo{person}{Rafael Goldstein}, {and} \bibinfo{person}{David Baraka}.} \bibinfo{year}{2021}\natexlab{}.
\newblock \showarticletitle{PrivacyFlash Pro: Automating Privacy Policy Generation for Mobile Apps.}. In \bibinfo{booktitle}{\emph{NDSS}}. \bibinfo{publisher}{Internet Society}, \bibinfo{address}{Reston, VA, USA}, \bibinfo{numpages}{18}~pages.
\newblock


\bibitem[Zimmeck et~al\mbox{.}(2019)]%
        {zimmeck2019maps}
\bibfield{author}{\bibinfo{person}{Sebastian Zimmeck}, \bibinfo{person}{Peter Story}, \bibinfo{person}{Daniel Smullen}, \bibinfo{person}{Abhilasha Ravichander}, \bibinfo{person}{Ziqi Wang}, \bibinfo{person}{Joel~R Reidenberg}, \bibinfo{person}{N~Cameron Russell}, {and} \bibinfo{person}{Norman Sadeh}.} \bibinfo{year}{2019}\natexlab{}.
\newblock \showarticletitle{MAPS: Scaling privacy compliance analysis to a million apps}.
\newblock \bibinfo{journal}{\emph{Proceedings on Privacy Enhancing Technologies (PoPETs)}} \bibinfo{volume}{2019}, \bibinfo{number}{3} (\bibinfo{year}{2019}), \bibinfo{pages}{66--86}.
\newblock


\bibitem[Zimmeck et~al\mbox{.}(2017)]%
        {zimmeck2017automated}
\bibfield{author}{\bibinfo{person}{Sebastian Zimmeck}, \bibinfo{person}{Ziqi Wang}, \bibinfo{person}{Lieyong Zou}, \bibinfo{person}{Roger Iyengar}, \bibinfo{person}{Bin Liu}, \bibinfo{person}{Florian Schaub}, \bibinfo{person}{Shomir Wilson}, \bibinfo{person}{Norman~M Sadeh}, \bibinfo{person}{Steven~M Bellovin}, {and} \bibinfo{person}{Joel~R Reidenberg}.} \bibinfo{year}{2017}\natexlab{}.
\newblock \showarticletitle{Automated Analysis of Privacy Requirements for Mobile Apps.}. In \bibinfo{booktitle}{\emph{NDSS}}. \bibinfo{publisher}{Internet Society}, \bibinfo{address}{Reston, VA, USA}, \bibinfo{numpages}{15}~pages.
\newblock


\end{thebibliography}
